\documentclass[10pt,journal,compsoc]{IEEEtran}
\ifCLASSOPTIONcompsoc
  \usepackage[nocompress]{cite}
\else
  \usepackage{cite}
\fi
\ifCLASSINFOpdf
\else
\fi
\usepackage{graphicx}
\usepackage{threeparttable}
\usepackage{float}
\usepackage{multirow}
\usepackage{caption}
\usepackage{subcaption}
\usepackage{booktabs}
\usepackage[below]{placeins}
\usepackage{makecell}
\usepackage{stfloats}
\usepackage{algorithm}  
\usepackage{algpseudocode}  
\usepackage{amsmath}  
\usepackage{supertabular}
\usepackage[figuresright]{rotating}
\usepackage{multirow}
\usepackage{graphicx}
\usepackage{threeparttable}
\usepackage[switch]{lineno}
\begin{document}
\title{Prediction of  Depression Severity Based on the Prosodic and Semantic Features with Bidirectional LSTM and Time Distributed CNN}

\author{Kaining~Mao,~\IEEEmembership{Student Member,~IEEE,}
        Wei Zhang,
        Deborah~Baofeng~Wang, 
        Ang~Li,
        Rongqi Jiao, 
        Yanhui Zhu, 
        Bin Wu, 
        Tiansheng Zheng, 
        Lei Qian, 
        Wei Lyu, 
        Minjie Ye,
        and Jie Chen,~\IEEEmembership{Fellow,~IEEE}

\IEEEcompsocitemizethanks{\IEEEcompsocthanksitem K. Mao, W. Zhang, A. Li are with the Department
of Electrical and Computer Engineering, University of Alberta, Edmonton,
Alberta, Canada.\protect\\
E-mail: \{mmao1, wzhang6, ang6\}@ualberta.ca

\IEEEcompsocthanksitem J. Chen is with the Department
of Electrical and Computer Engineering, University of Alberta, Edmonton,
Alberta, Canada.\protect\\
E-mail: jc65@ualberta.ca
\IEEEcompsocthanksitem M. Ye is with The Affiliated Kangning Hospital of Wenzhou Medical University, Wenzhou, Zhejiang, China. (Corresponding authors: M. Ye and J. Chen)\protect 
\IEEEcompsocthanksitem D. Wang, T. Zheng, Q. Lei, W. Lyu are with The Affiliated Kangning Hospital of Wenzhou Medical University, Wenzhou, Zhejiang, China\protect
\IEEEcompsocthanksitem R. Jiao, Y. Zhu, B. Wu are with the Wenzhou Medical University, Wenzhou, Zhejiang, China\protect\\
}% <-this % stops a space
% \thanks{Manuscript received April 19, 2005; revised August 26, 2015.}
}


\markboth{IEEE TRANSACTIONS ON Affective Computing}%
{Mao \MakeLowercase{\textit{et al.}}: Prediction of PHQ-8 Score Based on the Prosodic and Semantic Features with Bidirectional LSTM and Time Distributed CNN}



\IEEEtitleabstractindextext{%
\begin{abstract}
Depression is increasingly impacting individuals both physically and psychologically worldwide. It has become a global major public health problem and attracts attention from various research fields. Traditionally, the diagnosis of depression is formulated through semi-structured interviews and supplementary questionnaires, which makes the diagnosis heavily relying on physicians' experience and is subject to bias. However, since the pathogenic mechanism of depression is still under investigation, it is difficult for physicians to diagnose and treat, especially in the early clinical stage. As smart devices and artificial intelligence advance rapidly, understanding how depression associates with daily behaviors can be beneficial for the early stage depression diagnosis, which reduces labor costs and the likelihood of clinical mistakes as well as physicians bias. Furthermore, mental health monitoring and cloud-based remote diagnosis can be implemented through an automated depression diagnosis system. In this article, we propose an attention-based multimodality speech and text representation for depression prediction. Our model is trained to estimate the depression severity of participants using the Distress Analysis Interview Corpus-Wizard of Oz (DAIC-WOZ) dataset. For the audio modality, we use the collaborative voice analysis repository (COVAREP) features provided by the dataset and employ a Bidirectional Long Short-Term Memory Network (Bi-LSTM) followed by a Time-distributed Convolutional Neural Network (T-CNN). For the text modality, we use global vectors for word representation (GloVe) to perform word embeddings and the embeddings are fed into the Bi-LSTM network. Results show that both audio and text models perform well on the depression severity estimation task, with best sequence level $F_1$ score of 0.9870 and patient-level $F_1$ score of 0.9074 for the audio model over five classes (healthy, mild, moderate, moderately severe, and severe), as well as sequence level $F_1$ score of 0.9709 and patient-level $F_1$ score of 0.9245 for the text model over five classes. Results are similar for the multimodality fused model, with the highest $F_1$ score of 0.9580 on the patient-level depression detection task over five classes. Experiments show statistically significant improvements over previous works.
\end{abstract}

}

\maketitle

% is not selected <OR> if conference mode is selected - because compsoc
% conference papers position the abstract like regular (non-compsoc)
\IEEEdisplaynontitleabstractindextext

% \fi
\IEEEpeerreviewmaketitle

\ifCLASSOPTIONcompsoc
\IEEEraisesectionheading{\section{Introduction}\label{sec:introduction}}
\else
\section{Introduction}
\label{sec:introduction}
\fi

\IEEEPARstart{M}{ental} health disorder, such as depression, is considered one of the major challenges facing global society. During the COVID-19 pandemic, the prevalence of depression and anxiety is exacerbated in the general population\cite{mazza2020anxiety,salari2020prevalence,barzilay2020resilience,nguyen2020people}. By 2030, depression will be the second major cause of disability worldwide and thus it can impose a heavy healthcare burden globally\cite{mathers2006projections}. It is estimated that the average cost of treating depression in 2010 is 24,000 € per patient and the total cost can be as high as €92 billion in Europe \cite{olesen2012economic}. In the United States, depression causes an estimated loss of \$44 billion, due to the absence or low working efficiency \cite{stewart2003cost}. According to a report from the World Health Organization (WHO), over 264 million people of all ages suffer from depression in 2017 \cite{james2018global}. Nearly 50\% of people with depression worldwide have difficulty receiving therapy \cite{rahman2008cognitive}.  Suicide is one of the severe results of depression, and the WHO reports that the number of people who passed away due to suicide is over 800,000 every year\cite{world2014preventing}. The number of attempted suicide is more frequent, possibly no less than 20 times that of those who died by suicide \cite{world2014preventing}. Patients with depression are more apt to generate suicide thoughts \cite{hawton2013risk},\cite{lepine2011increasing}. It is estimated that more than 50\% of people who died by suicide meet clinical criteria of depression\cite{joiner2005psychology},\cite{mcgirr2007examination}. However, often the symptoms of depression are not displayed directly. Many individuals often express their sadness and hopelessness but without depression, whereas patients are usually reluctant to report their conditions and receive treatment\cite{rodrigues2014impact}. For instance, many people with depression ignore or refuse to admit their emotional instability and physical health conditions. The reason is that depression is a stigmatized disease, resulting in the depressive population hiding or camouflaging their symptoms. Traditionally, a semi-structured clinical interview based on Diagnostic Statistical Manual (DSM) criteria is the standard protocol for depression diagnosis\cite{first2004structured} with self-test questionnaires such as the Patient Health Questionnaire Depression Scale (PHQ)\cite{kroenke2001phq}, Beck’s Depression Inventory (BDI)\cite{beck1961inventory} and Montgomery-Asberg Depression Rating Scale (MADRS)\cite{montgomery1977new}. The PHQ-8 is an assessment form created to examine the existence of the core depression symptoms, such as fatigue and anxiety. The PHQ-8 scale shows high sensitivity and specificity for diagnosing depression and other mental disorders among patients with different languages and cultures\cite{rude2004language}. These methods play a key role in diagnosing depression, but the results are subject to physicians' experience. Previous articles argued that these clinical criteria, such as DSM and BDI,  are not reliable enough\cite{bagby2004hamilton}. Diagnosis of depression is not the same as other medical conditions since gold standards for mental disorders do not exist currently, which raises the likelihood of misdiagnosis and finally leads to unexpected results \cite{kraemer2012dsm, kapur2012has,aragon2019detecting}. However, most depressed people do not have access to qualified psychological treatment due to economic conditions (low-/mid-income population) or living constraints (in rural regions)\cite{reece2017forecasting}. Currently, Schuller et al proved that infrastructure such as a high-speed network and smartphones with high-performance computational units can provide support for continuous monitoring of the psycho-emotional state for a long period\cite{han2021deep}. Therefore, it will be beneficial to develop a low-cost screening technique that can be deployed in communities and operated by people without special training. Early-stage mental disorder screening is also crucial for policymakers and security agencies because someone with mental health disorder could behave adversely to other innocent people, such as massive shootings which are attributed to mental health disorders\cite{joyal2007major}. Recently, the development of Artificial Intelligence (AI) shows its great potential in healthcare\cite{cummins2020artificial}. In the cyber world, we live now, it is very common to share personal information, concerns through the Internet, especially after the rise of social media. This raises an opportunity since the contents on social media increase the likelihood to detect potential depression patients from a large population. The automated depression diagnosis has been studied from many different aspects, such as collecting and analyzing data, induction and representation of emotions, and predicting depression based on different modalities.
 In this paper, we propose a multimodality automated depression diagnosis system with prosodic and semantic features to predict the depression levels with the combination of Bi-LSTM and T-CNN models. To the best of our knowledge, it is the first time that time-distributed CNN is adopted to further extract the temporal information from the output of the LSTM encoder. Additionally, our proposed model does not have a strict limitation of input duration, regardless of the number of frames, as long as the number of features meets our specification, our model can always provide a patient-independent depression prediction. The prediction is based on a specific text or audio feature sequence. Given a specific participant with an audio/text feature sequence of arbitrary length, our model provides a series of estimations of depression severity based on the audio/text feature. The set of predictions can be merged through a major voting algorithm so that the final output of our model is a patient-level depression severity prediction. This mitigates the problem that the audio/text feature sequences are required to be the same in length in previous articles. LSTM performs well in learning temporal information because of its recurrent structure. The bidirectional LSTM model is used to learn long-term bidirectional dependencies in the audio and text feature sequences because it has been proved to perform better than a unidirectional LSTM model. The convolutional neural network (CNN) is a popular network architecture to learn the spatial features of data. A time-distributed CNN architecture is obtained by having multiple CNN layers for Bi-LSTM output features at each timestep. Given the complementary advantage of CNN and LSTM, the hybrid model of LSTM and T-CNN works well in learning the spatiotemporal sequence. The best patient-independent $F_1$ score of the audio and text model is 0.9870 and 0.9709, respectively, on the test partition of the DAIC-WOZ dataset. The fused multimodality model achieved the best $F_1$ score of 0.9580 on the test partition of the DAIC-WOZ dataset.

\begin{figure*}[!t]
\centering
\scalebox{0.9}{\includegraphics[width=\textwidth]{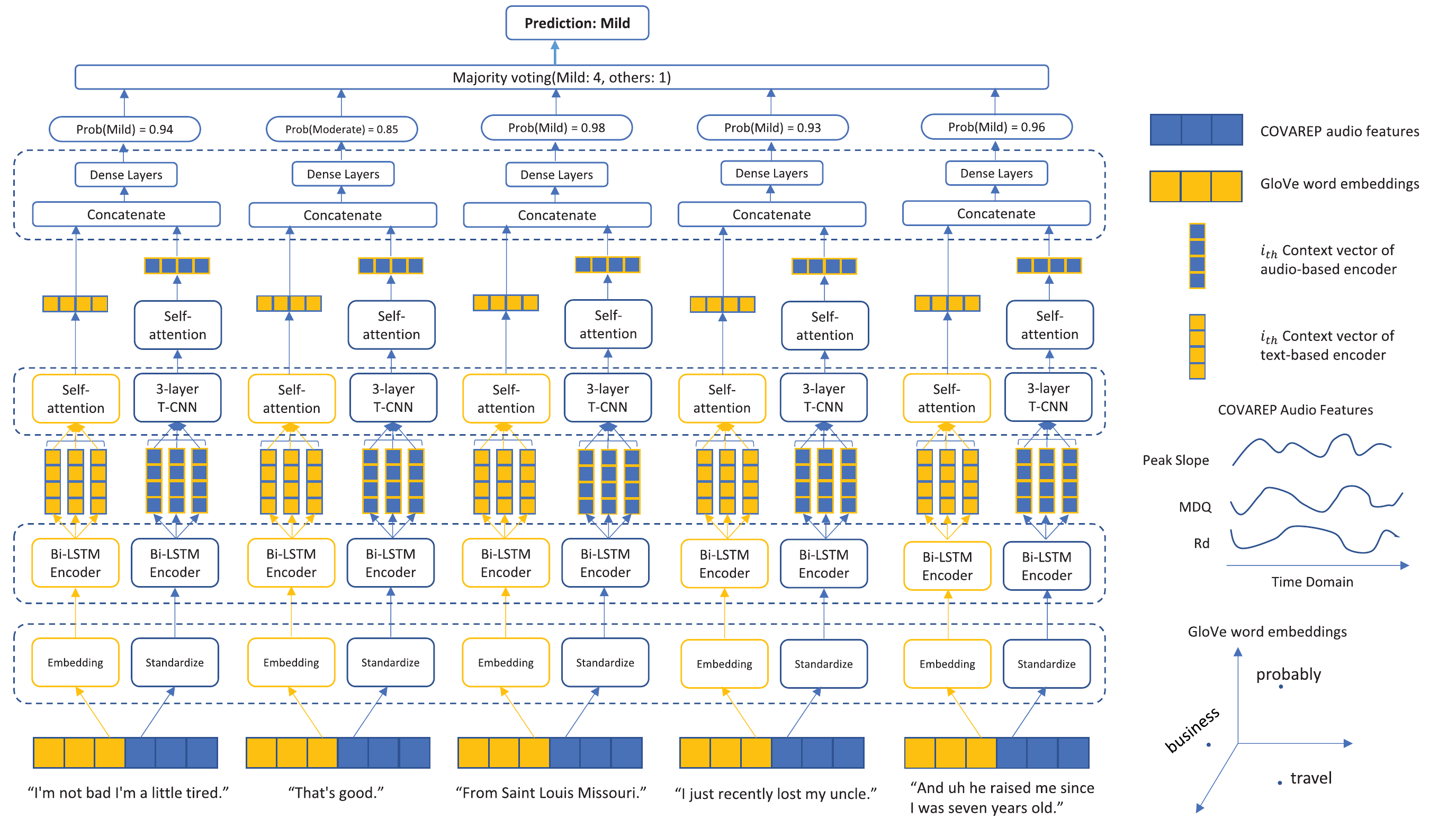}}
\caption{Block diagram of our proposed multimodality depression level prediction algorithm given a specific example. Audio features are fed into the network through the input layer. After batch normalization, the input data is fed into the Bi-LSTM and time-distributed CNN block. In this proposed design, we have five time-distributed CNN blocks followed by a single-layer Bi-LSTM. The detailed architecture of each block is illustrated and explained in the remainder of this paper.}
\label{fig:scheme_whole}
\end{figure*}


\section{RELATED WORK}
\subsection{Depression Severity Prediction Based on Text}
Text, especially non-verbal cues which are not expressed directly in dialogue has gained remarkable popularity in depression prediction and sentiment analysis due to two reasons. Firstly, psychiatrists observe speech attributes such as less variation in speech production during an interview, which are commonly used biomarkers in depression diagnosis\cite{aragon2019detecting,morales2018linguistically,pirina2018identifying,yates2017depression,ray2019multi}. Secondly, the text transcription is an explicit signal to record, which makes interview transcripts one of the best candidates for distress analysis tasks \cite{rude2004language},\cite{resnik2013using}. The prediction of depression severity is based on the hypothesis that mental disorder causes some accessible and observable differences affecting how verbal content is produced \cite{scherer1986vocal,kroenke2010patient}. Previously, a problem in searching for the relationship between depression and semantic features is the difficulty in collecting qualified and sufficient data. With the advancement of social network usage, a large amount of text data inflow gives an opportunity to researchers to analyze distress state from text \cite{aragon2019detecting,pirina2018identifying,yates2017depression,hiraga2017predicting}. Coppersmith et al. seminally proposed to acquire a qualified dataset via social network platforms, which solves the problem of data insufficiency\cite{coppersmith2015clpsych}. However, those colloquial languages, such as abbreviated words, popular slang, etc., make data preprocessing very difficult. Additionally, people are more likely to publish negative content on social media because they are anonymous. This indicates that although someone without any mental disorders is still likely to publish many negative posts for some periods. The effectiveness of social media posts should be fully investigated before being widely used in automated depression diagnosis because the quality of the training data affects the performance of the classifier. We also disagree that collecting data from the Internet is an effective strategy because those posts not related to depression are likely to be dominating factors compared with those depression-sensitive posts. To solve the problem of patients concealing their thoughts and emotional states,  Scherer et al. proposed to collect the dialogue during the screening interview led by clinicians \cite{scherer2014automatic}. This data collection strategy has several advantages. Firstly, the questions are specifically designed by psychologists, and the efficiency of the question is better than user-generated content on social media. Before starting the interview, the participants are required to complete a questionnaire like PHQ-8 or BDI. After the interview, the clinician determines the depression severity of the participant based on the response of the patient during the interview and the questionnaire. Overall, these studies highlight the need for reliable corpora for speech-based depression prediction.

\subsection{Depression Severity Prediction Based on Voice}
In this section, we overview the application of prosodic and acoustic features in predicting depression. The relationship between depression and voice change has been well-studied \cite{cannizzaro2004voice,scherer2013investigating}. The earliest research on depressive voices can be traced back to the 1920s. The father of modern psychiatry, Emil Kraepelin, characterized the voice to be depressive as “low voice, slowly, hesitatingly, monotonously, sometimes shuttering, whispering, try several times before they bring out a word, become mute in the middle of a sentence \cite{kraepelin1921manic}.” To train the audio model, the first thing is to extract audio features from the raw audio recordings. Feature extraction is the preprocessing technique that converts the original audio into more abstract, dense vectors. Cummins et al. pointed out several critical properties for a perfect feature to detect depression or other mental disorders\cite{cummins2015review}. The most important property is that the feature should represent some recurring and noticeable effects caused by depression. The feature must also manifest large cross-label variability but small inner-label variability. Furthermore, the feature should be robust to environmental noise if it is intended to be used in the automated depression diagnosis system.
Many previous works adopted a Support Vector Machine (SVM) and Gaussian Mixture Model (GMM) \cite{williamson2013vocal, alghowinem2013comparative}. They are two popular machine learning techniques and are robust to overfitting. Much of the available literature has attempted to use the combination of prosodic and glottal features to train the classifier \cite{cummins2013spectro, moore2007critical, cohn2009detecting, trevino2011phonologically, scherer2013investigating}. As for the Mel-Cepstral Coefficients, it is reported that the low-order MFCCs perform much better than high-order MFCCs for emotion prediction or some para-linguistic analysis tasks\cite{bone2014robust}. Additionally, except for these low-level audio features, some researchers proposed to use pre-trained convolutional neural networks such as VGG-16 to extract high-level features in a frequency spectrogram\cite{ray2019multi}. However, the effectiveness of this deep frequency spectrum feature is questionable. Although the CNN model outperforms other traditional models in Computer Vision, the frequency spectrogram is different from other images. The CNN is spatial invariant because it applies a group of identical transformations to different regions of an image\cite{gratch2014distress}. The frequency spectrogram consists of an X-axis denoting the frequency and the Y-axis as the intensity of the frequent component. The position of a component in the frequency spectrogram matters, but the components in ordinary images are less sensitive to the position. Regarding this concern, we do not adopt this deep frequency spectrum feature extraction method. Instead, our model utilizes the low-level audio features mentioned above. Together, these studies indicate that we should consider the combination of audio features as the input for training the depression prediction model.

\section{DATASET AND METHODS}
\label{sec:dnm}
In this section, we briefly introduce the preliminary material we used for developing the audio model, text model, and multimodality model. We also discuss the dataset and framework for training and evaluating our proposed model.
\subsection{Distress Analysis Interview Corpus-Wizard of Oz (DAIC-WOZ)}
In this paper, we adopted the Distress Analysis Interview Corpus-Wizard-of-oz (DAIC-WOZ) dataset for training and testing \cite{gratch2014distress}. The corpus consists of 189 recorded clinical interviews and transcripts as well as facial features from 189 subjects. The audio recordings were taken of semi-structured interviews between the participants and a virtual interviewer called Ellie, an animated role controlled by a human interviewer. The average audio duration of 189 subjects is 974 seconds. Subjects were solicited from the Greater Los Angeles Metropolitan region from two different populations. One was from civilians; the other was from veterans of the U.S armed forces. Subjects were characterized as depression, Post-Traumatic Stress Disorder (PTSD), and anxiety based on the self-report questionnaire during the data collection\cite{gratch2014distress}. Only the interview recordings of the depression group were released for academic purposes. The gender distribution over all five groups as well as the dataset partition is shown in Table. \ref{tab:gender}. In the training set, there are 44 female subjects (27 without significant depression symptoms, 17 with depression symptoms) and 63 male subjects (49 without significant depression symptoms, 14 with depression symptoms). In the validation set, there are 19 female subjects (12 without significant depression symptoms, 7 with depression symptoms) and 16 male subjects (11 without significant depression symptoms, 5 with depression symptoms). In the test set, there are 24 female subjects (17 without significant depression symptoms, 7 with depression symptoms) and 23 male subjects (16 without significant depression symptoms, 7 with depression symptoms). All interviews were transcribed verbatim into English. The interviews lasted from 5 to 20 minutes involving three phases: it started with neutral questions, which aimed to ensure subjects being able to calm down; the interview then proceeded into a targeted phase, and the questions asked by the interviewer were more related to the symptoms of depression and PTSD. Finally, the interview terminated with the annealing phase, which assisted the participants to get rid of the distressed state. The PHQ-8, ranging from 0 to 24, determines the severity of the mental disorder. Subjects were divided into five groups: healthy (PHQ-8\textless5), mild (5\textless PHQ-8\textless 10), moderate (10\textless PHQ-8\textless15), moderately severe (15\textless PHQ-8 \textless20), and severe (PHQ-8 \textgreater20)\cite{kroenke2009phq}. Table. \ref{table:showcase} shows a sample transcript in the DAIC-WOZ dataset, which contains four fields: beginning and end timestamp of the utterance, the speaker ID, and sentence content. Due to space limitation, Fig. \ref{fig:showcase} below illustrates the distribution of the first four audio features provided with the DAIC-WOZ dataset with the significant intra-subject variance. In the remaining part of this paper, the training, validation and test set are split by the instruction from the DAIC-WOZ dataset independently, which ensures all the subjects only appear in one of the above partitions.
\begin{table}[htb]
\caption{The Showcase of a Participant's Transcript}
\centering
\label{table:showcase}
\scalebox{0.8}{
\begin{tabular}{cccc}
\hline
\textbf{Start time} & \textbf{Stop time} & \textbf{Speaker}     & \textbf{Utterance}                                \\ \hline
87.322     & 89.592    & Ellie       & So how are you doing today?              \\
89.71      & 91.93     & Participant & I'm not bad I'm a little tired but okay. \\
92.945     & 93.585    & Ellie       & That's good.                             \\
94.257     & 95.577    & Ellie       & Where are you from originally?           \\
95.78      & 97.14     & Participant & Uh from Saint Louis, Missouri.            \\ \hline
\end{tabular}
}
\end{table}

\begin{figure}[htb]
\centering
\scalebox{0.8}{\includegraphics[width=0.5\textwidth]{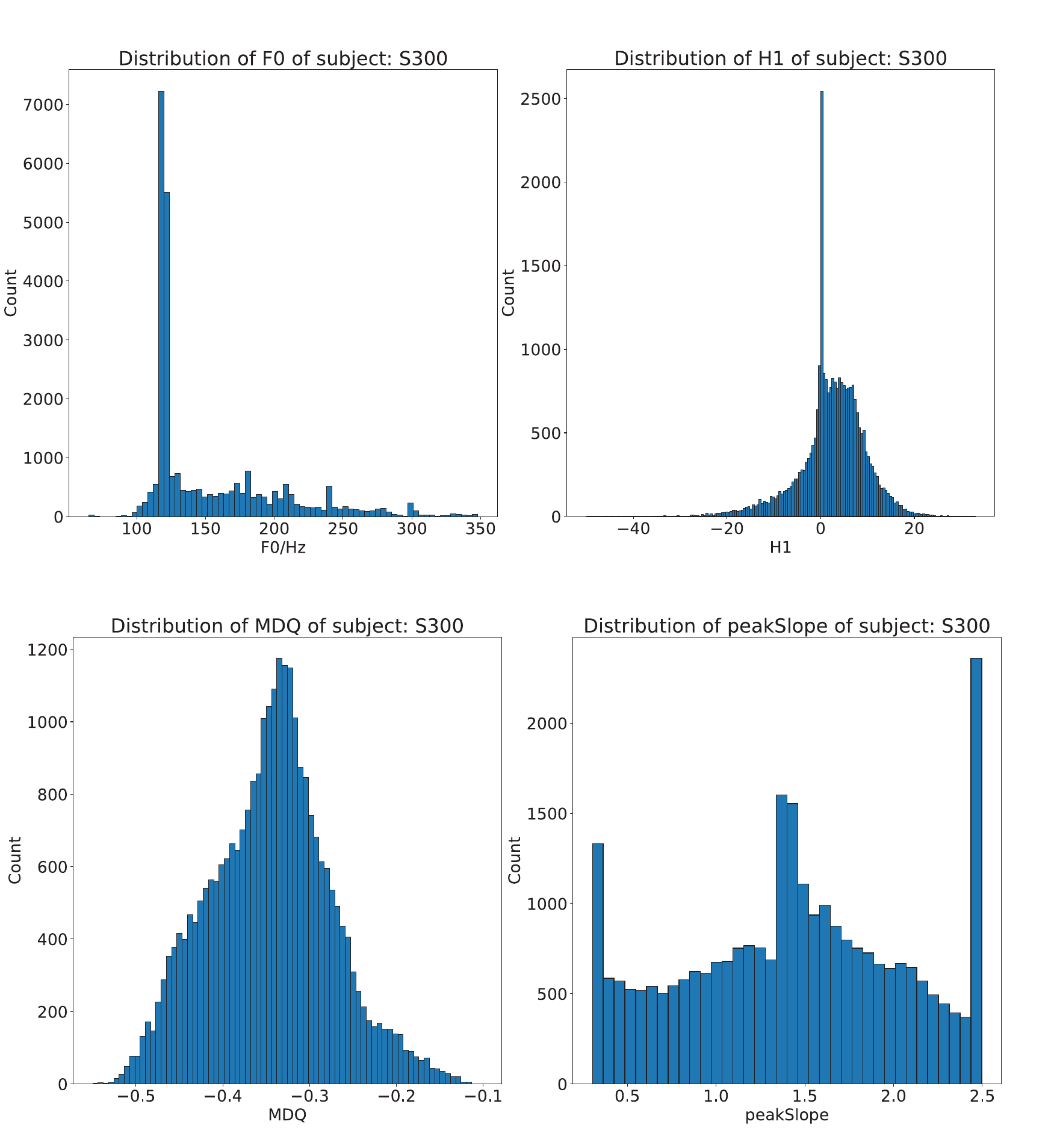}}
\caption{The distribution of a subset of audio features within the dataset}
\label{fig:showcase}
\end{figure}

\subsection{Recurrent Neural Network and Long-Short Term Memory }
\begin{figure}[!t]
\centering
\includegraphics[width=3.5in]{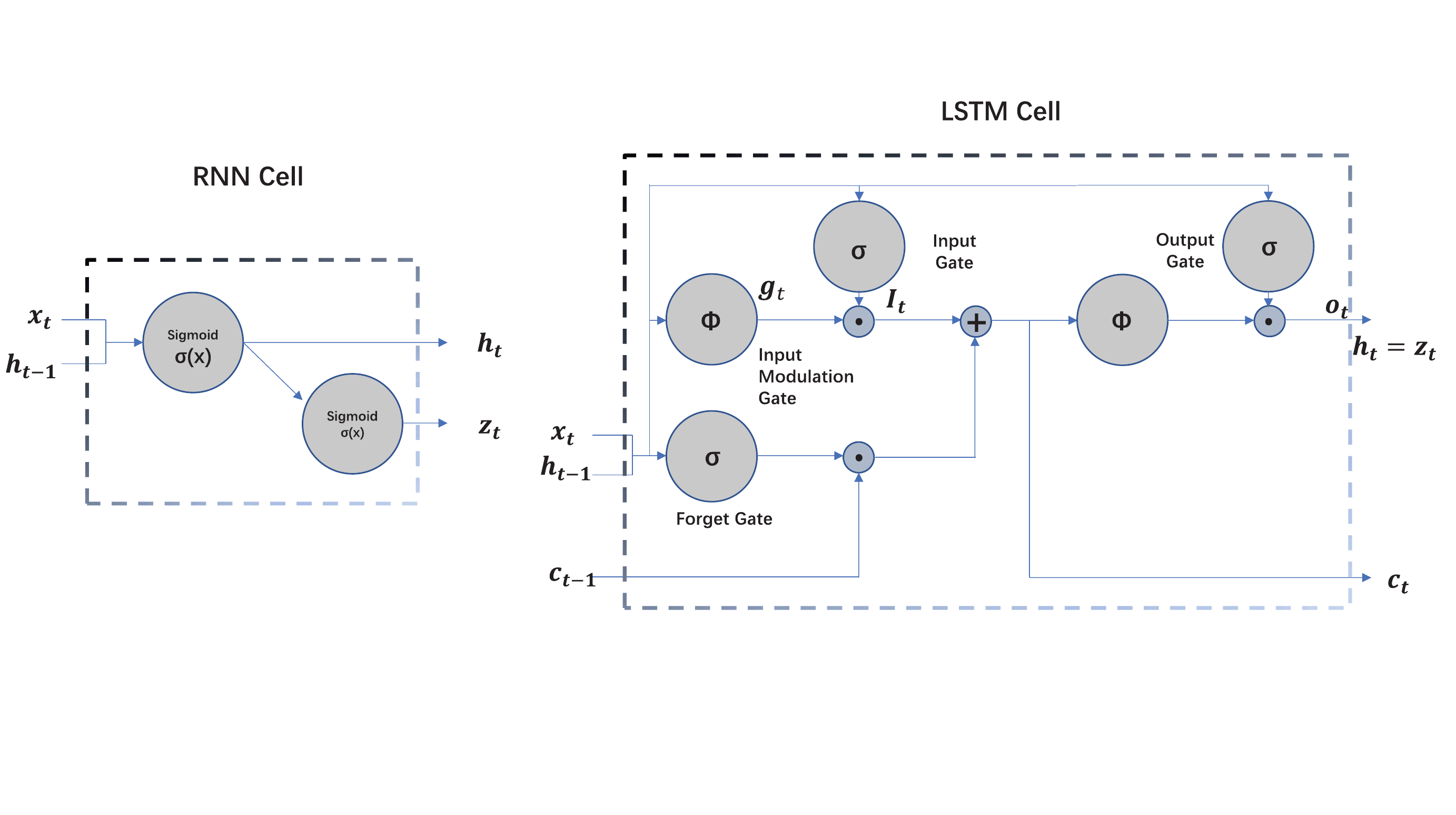}
\caption{A schematic of an RNN cell (left) and an LSTM cell (right) used in this paper.}
\label{fig:schem_rnn_lstm}
\end{figure}

A recurrent neural network (RNN) is a deep learning architecture that outputs a time sequence. The input of the neural network is transformed into hidden states at different time steps. Given an input vector $x_t$, the intermediate variables in the network are computed iteratively, from ($h_1$, $z_1$), ($h_2$, $z_2$) to ($h_t$, $z_t$), where $h_t$ and $z_t$ are the hidden state and output of the RNN cell, respectively. The traditional RNN performs well on some machine learning tasks, such as voice recognition\cite{ray2019multi}. However, the gradient vanishing/exploding problem during the backpropagation limits the depth of the RNN. To solve this problem, Hochreiter et al. proposed the LSTM, which stands for Long-Short Term Memory\cite{crawford2019spatially}. LSTM can determine when to “forget” some previous information and update the hidden state during the training phase by combining different “gates” in the LSTM cell. The traditional RNN and the LSTM cell are illustrated in Fig.\ref{fig:schem_rnn_lstm}. 
Compared with the traditional RNN cell, the LSTM cell includes some special components such as the input gate, forget gate, output gate, input modulation gate, and the memory cell. The $i_t$ and $f_t$ determine whether the previous information should be memorized or forgotten. Similarly, the output gate determines how much information in the cell memory can be transferred to the hidden state. These gates enhance the performance of LSTM on time series-related tasks and make it possible to train a deeper network. The hidden state of the previous layer can be fed into the following layers to construct a deeper network, which improves the capability of LSTM to deal with more complicated time series.
From the probabilistic perspective, automated depression diagnosis is to find a correct severity sequence $y$ that maximizes the conditional probability of $y$ given an input feature sequence (i.e., audio/text feature). Our proposed framework, based on an RNN encoder-decoder, learns to predict depression severity given a sequence of audio and text features. In the encoder neural network, an encoder reads and projects the input feature sequence $X=\left(x_1,x_2,\ldots,x_T\right)$ into a context vector $c$, which is given by:
\begin{equation}
    c=q\left(h_1,\ldots,h_T\right)
\end{equation}
\begin{equation}
    h_t=f\left(x_t,h_{t-1}\right)
\end{equation}
where $h_t$ is a hidden state at time $t$, $c$ is a vector computed from a sequence of hidden states. $f$ and $q$ are nonlinear functions. The decoder neural network is trained to predict the depression severity given the context vector and the input feature at time $t$. The probability of depression severity is given by:
\begin{equation}
    p\left(y\right)=\prod_{t=1}^{T}{p\left(y_t\mid\left\{x_1,\cdots,x_{t-1}\right\},c\right)}
\end{equation}
where $y=\left(y_1,\cdots,y_T\right)$, and each term of the conditional probability is given by:
\begin{equation}
    p\left(y_t\mid\left\{x_1,\cdots,x_{t-1}\right\},c\right)=g\left(x_{t-1},s_t,c\right)
\end{equation}
where $g$ is a nonlinear function, $s_t$ is the hidden state of the RNN.

\subsection{Attention Mechanism}
Most of the proposed deep learning-based depression prediction models are a member of a family of encoder-decoders, with an encoder for high-level representation of original input audio or text features. The encoder network reads and encodes the variable-length input audio/text features into a fixed-length vector. A decoder then decodes the fixed-length vector and outputs a probability matrix from the encoded fixed-length vector. The cascade model, which consists of an encoder and decoder, is optimized simultaneously to maximize the probability of a correct depression severity given an original audio/text feature sequence.

A shortcoming of this encoder-decoder architecture is that the encoder network has to compress all the depression-sensitive information into a fixed-length vector. In the scenario of extremely long input sequences, this may make it challenging for the encoder network to encode necessary information into the fixed-length vector, especially during testing, when the length of the input sequence for testing is longer than the length of sequence for training. To overcome this drawback, we adopted an attention mechanism that allows the model to select a subset of encoded vectors adaptively while decoding the high-level representation. Each time the decoder makes an inference on depression severity, it goes through the encoded input sequence and works out the most depression-sensitive information. The most important feature of the attention mechanism is that it does not rely on a single fixed-length vector. The model can select a subset of encoded high-level representation adaptively during training, which frees the encoder network from compressing all necessary depression-related information, no matter how long the original sequence is, into a fixed-length vector. This improves the performance of our model, especially the performance coping with long sequences. 

With the attention mechanism, we can compute the weighted context vector with RNN output hidden states. The depression conditional probability of time step t is given by:
\begin{equation}
    p\left(y_t\mid\left\{x_1,\cdots,x_{t-1}\right\},c\right)=g\left(x_{t-1},s_t,c_t\right)
\end{equation}
Where $s_t$ is the RNN hidden state for time $t$, which is given by:
\begin{equation}
    s_t=f\left(s_{t-1},x_{t-1},c_t\right)
\end{equation}

Unlike the traditional encoder-decoder framework, the depression conditional probability is not only conditioned on a uniform context vector $c$ but a distinct vector $c_t$ for each timestep. The context vector is given by a sequence of RNN hidden states, which are the output of the encoder neural network. A hidden state at time step $t$ contains all information about the input feature sequence prior to time step $t$, with an emphasis on the part around the entries at time step $t$. The context vector is given by:
\begin{equation}
    c_t=\sum_{j=1}^{T}{\alpha_{ij}h_j}
\end{equation}
The coefficients $\alpha_{ij}$ for each hidden state is determined by:
\begin{equation}
    \alpha_{ij}=\frac{\exp{\left(e_{ij}\right)}}{\sum_{k=1}^{T}\exp{\left(e_{ik}\right)}}
\end{equation}
where $e_{ij}$ is given by:
\begin{equation}
    e_{ij}=a\left(s_{i-1},h_j\right)
\end{equation}
$a(x)$ is a score function that evaluates how well the inputs around the entries at time $j$ and the output of RNN at time $(i-1)$ match. The score function $a(x)$ is a distinct layer that is simultaneously trained with all other layers of the proposed model. The probability $\alpha_{ij}$ describes the importance of the hidden state $h_j$ regarding the previous hidden state $s_{i-1}$ during calculation of $s_i$. This allows the decoder itself to determine which part of the input sequence should be focused on. With the attention mechanism, we alleviate the burden of compressing the input sequence, regardless of its original length, to a fixed-length vector. Therefore, with the attention mechanism, the correlation in the context vector can be propagated through the network, which allows the decoder to selectively retrieve those depression-related hidden states.
\begin{figure}[htb]
        \centering
        \begin{subfigure}[b]{0.475\textwidth}
            \centering
            \includegraphics[width=2.5in]{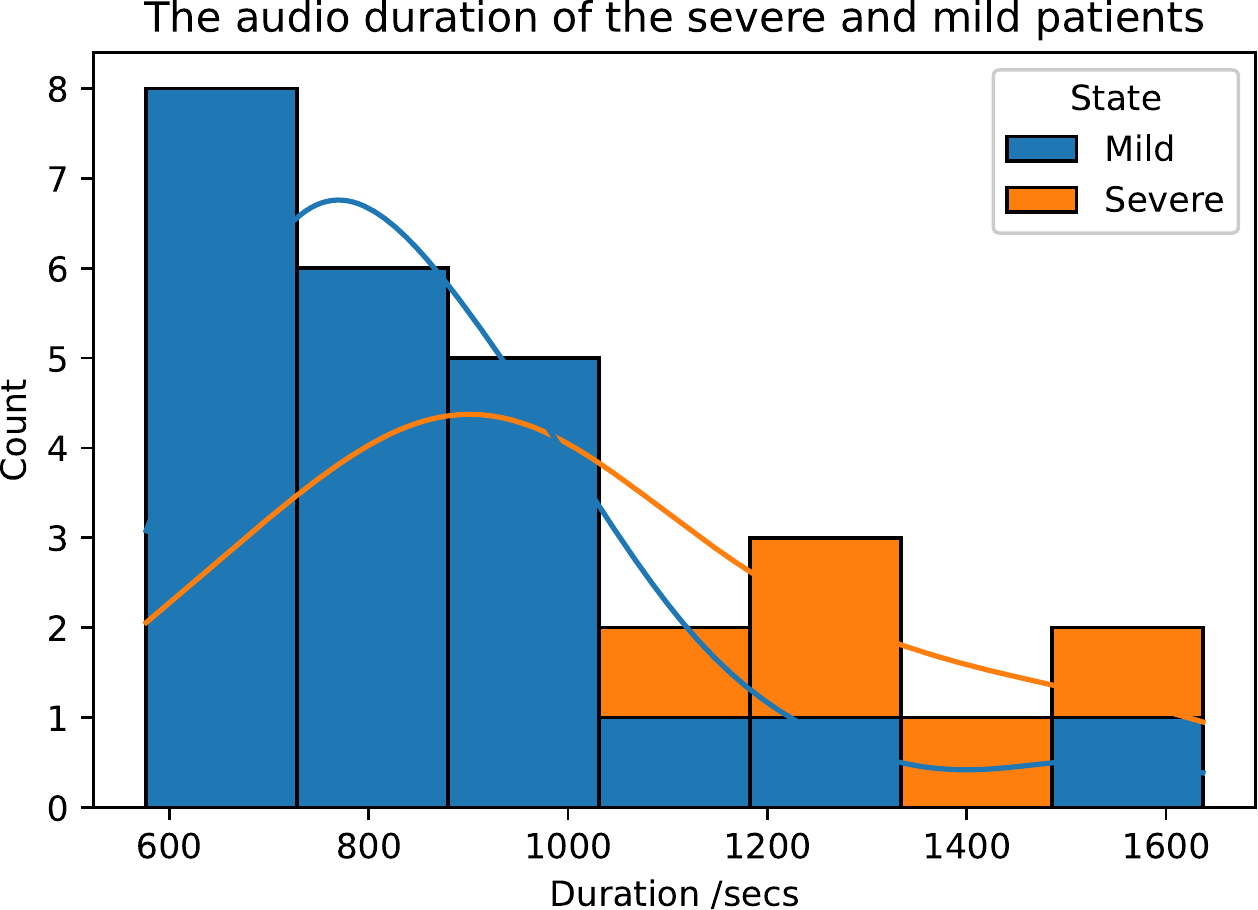}
            \caption[Confusion matrix of the audio model with time step = 16]%
            {{\small }}    
            \label{figure:audio_duration}
        \end{subfigure}
        \hfill
        \begin{subfigure}[b]{0.475\textwidth}  
            \centering 
            \includegraphics[width=2.5in]{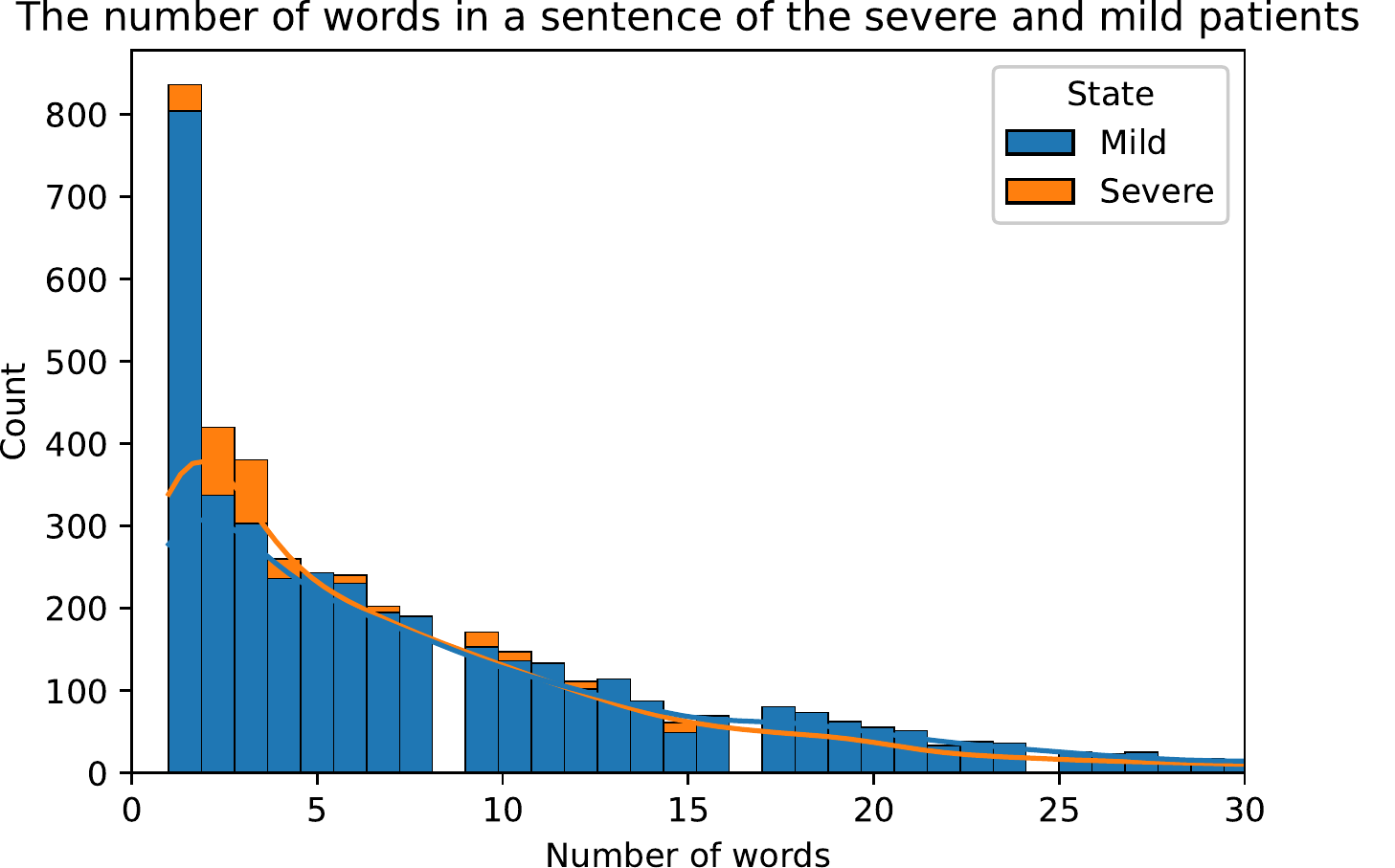}
            \caption[]%
            {{\small }}    
            \label{figure:sentence_length}
        \end{subfigure}
        \caption[]
        {\small The histograms of the audio duration and sentence length of control and experiment groups. (\subref{figure:audio_duration}) The audio duration of the control and experiment groups. (\subref{figure:sentence_length}) The sentence length of the control and experiment groups.} 
        \label{figure:audio_sentence_length}
\end{figure}

\begin{table}[htb]
\centering
\caption{RESULT OF THE T-TEST OF THE CONTROL AND EXPERIMENT GROUP}
\label{table:result_t_test}
\scalebox{0.75}{
\begin{tabular}{lcccc}
\toprule
        & \multicolumn{2}{c}{\textbf{Sentence length}} & \multicolumn{2}{c}{\textbf{Audio duration}}    \\
        \midrule
        & \textbf{Control}          & \textbf{Experiment}       & \textbf{Control}            & \textbf{Experiment}       \\
        & 8.7854±8.9475    & 7.3717±7.2975    & 951.3711±266.6010 & 997.8773±290.1901 \\
\textbf{p-value} & \multicolumn{2}{c}{$3.2397 \times 10^{-14}$}          & \multicolumn{2}{c}{0.0952}     \\
\bottomrule
\end{tabular}
}
\end{table}

\begin{table*}[htb]
\centering
\caption{Gender Distribution Over All Groups and Dataset Partitions}
\label{tab:gender}
\resizebox{\textwidth}{!}{%
\begin{tabular}{ccccccc}
\toprule
 & \multicolumn{2}{c}{Training set} & \multicolumn{2}{c}{Validation set} & \multicolumn{2}{c}{Test set} \\ \midrule
ID & \multicolumn{2}{l}{\begin{tabular}[c]{@{}l@{}}303, 304, 305, 310, 312, 313, 315, 316, 317, 318, 319, 320, 321, 322, \\ 324, 325, 326, 327, 328, 330, 333, 336, 338, 339, 340, 341, 343, 344, \\ 345, 347, 348, 350, 351, 352, 353, 355, 356, 357, 358, 360, 362, 363, \\ 364, 366, 368, 369, 370, 371, 372, 374, 375, 376, 379, 380, 383, 385, \\ 386, 391, 392, 393, 397, 400, 401, 402, 409, 412, 414, 415, 416, 419, \\ 423, 425, 426, 427, 428, 429, 430, 433, 434, 437, 441, 443, 444, 445, \\ 446, 447, 448, 449, 454, 455, 456, 457, 459, 463, 464, 468, 471, 473, \\ 474, 475, 478, 479, 485, 486, 487, 488, 491\end{tabular}} & \multicolumn{2}{l}{\begin{tabular}[c]{@{}l@{}}302, 307, 331, 335, 346, 367, 377, 381, 382, 388, 389, 390, 395, 403, \\ 404, 406, 413, 417, 418, 420, 422, 436, 439, 440, 451, 458, 472, 476, \\ 477, 482, 483, 484, 489, 490, 492\end{tabular}} & \multicolumn{2}{l}{\begin{tabular}[c]{@{}l@{}}300, 301, 306, 308, 309, 311, 314, 323, 329, 332, 334, 337, 349, 354, \\ 359, 361, 365, 373, 378, 384, 387, 396, 399, 405, 407, 408, 410, 411, \\ 421, 424, 431, 432, 435, 438, 442, 450, 452, 453, 461, 462, 465, 466, \\ 467, 469, 470, 480, 481\end{tabular}} \\ \bottomrule
\multicolumn{7}{c}{Dataset profile for depression level classifcation} \\ \midrule
 & Female & Male & Female & Male & Female & Male \\
\#Healthy & 7 & 9 & 2 & 3 & 3 & 2 \\
\#Mild & 12 & 25 & 7 & 6 & 9 & 11 \\
\#Moderate & 10 & 20 & 5 & 2 & 7 & 3 \\
\#Moderately severe & 10 & 5 & 2 & 2 & 1 & 4 \\
\#Severe & 5 & 4 & 3 & 3 & 4 & 3 \\

\bottomrule
\multicolumn{7}{c}{Dataset profile for depression detection} \\ \midrule
\#Subjects w/o significant symptom (PHQ-8$\leq$10) & 27 & 49 & 12 & 11 & 17 & 16 \\
\#Subjects w/ significant sympotom (PHQ-8$\textgreater$10) & 17 & 14 & 7 & 5 & 7 & 7 \\

\bottomrule
\end{tabular}%
}
\end{table*}

\subsection{Audio Features and Models}
\label{sec:audio}
In this paper, the audio features are extracted by COVAREP\cite{degottex2014covarep}, which can be divided into three categories: glottal flow features (NAQ, QOQ, H1-H2, PSP, MDQ, Peak slope, Rd), voice quality features ($F_{0}$, VUV), and spectral features (MCEP, HMPDM, HMPDD). Normalized Amplitude Quotient (NAQ) quantifies the time-based feature of the speaker by amplitude-domain measurements calculated from the glottal flow and its first derivative\cite{alku2002normalized, hacki1989klassifizierung}, Quasi Open Quotient (QOQ), which is a correlate of the open quotient (OQ) which involves the derivation of the quasi-open phase based on the amplitude of the glottal phase\cite{kane2013comparative, holmberg1995comparisons}, the amplitude difference of the first two harmonics of the differentiated glottal source spectrum (H1H2) \cite{alku1997parabolic}, Parabolic Spectral Parameter (PSP), which is based on the quantification of the spectral decay of the speaker\cite{alku1997parabolic}, and Maxima Dispersion Quotient (MDQ), which is designed to quantify the maxima dispersion as a result of phonation type moves towards a breathier phonation\cite{kane2013wavelet, huang2001spoken}. Spectral features consist of Mel-Cepstral Coefficients (MCEP0-24), which is a representation of the short-term power spectrum of a sound\cite{rodrigues2019multimodal}, harmonic model and phase distortion mean (HMPDM0-24) and deviation (HMPDD0-12). Thus, there are 74 audio features in total. Each subject is represented in the COVAREP features, $X_i \in R^{T \times F}$ where $T$ denotes the time dimension, which is proportional to the duration of the audio. Each 10 milliseconds frame of audio was transformed into an audio feature vector. $F$ denotes the number of features COVAREP extracted for each frame. Among the 74 audio features, the entry “VUV” indicates whether the audio features are extracted from the audible or silent part of the original interview recording. Only those audio features where "VUV" is 1 can be the input to the following models. Among all the 189 subjects in the dataset, audio features are in an average of 35850 frames (rows) and a standard deviation of 15791 frames (rows). For each subject, we concatenated a constant number of audio feature frames into a set of successively retrieved audio feature sequences, which were used to represent this subject. The shape of the input tensor is thus (\#samples, \#frames, 73). The field "VUV" is always 1 in the input tensor so it is dropped, which results in the final input tensor shape as 73. \par
Audio models with different configurations for depression assessments are introduced as follows. The input to these models is the previously mentioned audio feature sequences, the output of these models is the prediction of the depression severity given an audio feature sequence. The first audio model is a simple one that consists of the LSTM and fully connected layers. The LSTM served as a feature extractor and the following fully connected layers made the prediction based on the output of the LSTM. Then, we introduce our proposed model that consisted of the Bi-LSTM and T-CNN and they were evaluated for the prediction of depression severity.
\subsubsection{Traditional LSTM-based Model}
\label{sec:lstm_text}
Our first audio model comprises of single-layer Long-Short Term Memory (LSTM) network and fully connected layers. LSTM network was obtained using an LSTM layer containing 73 hidden units, connected to a fully connected layer. To avoid overfitting, the dropout was applied to the recurrent input signal on the LSTM units and between fully-connected layers with the dropout rate of 0.2. The time step is equal to the constant "\#frames" and there were 73 features in each timestep. In this model, only the hidden state at the last time step was fed into the following fully connected layers, with 128 and 64 hidden units. The output of the fully connected layer was then fed into a batch normalization layer and flattened into a 1D tensor. The flattened tensor was fed into a fully connected layer with 5 hidden units, where the SoftMax activation function transformed the unnormalized output of each neuron into the probabilities of five severities. An Adam optimizer was adopted for the training, the initial learning rate was set to be 0.001, $\beta_1$=0.9, $\beta_2$=0.999 and the epsilon was $10^{-7}$. A callback function monitored the validation loss and terminated the training if the validation loss did not decrease after five epochs. A loss function of cross-entropy was applied.
\subsubsection{Hybrid of Bi-LSTM and T-CNN Model}
\begin{figure}[htb]
    \centering
    \scalebox{0.7}{\includegraphics[width = 0.5\textwidth]{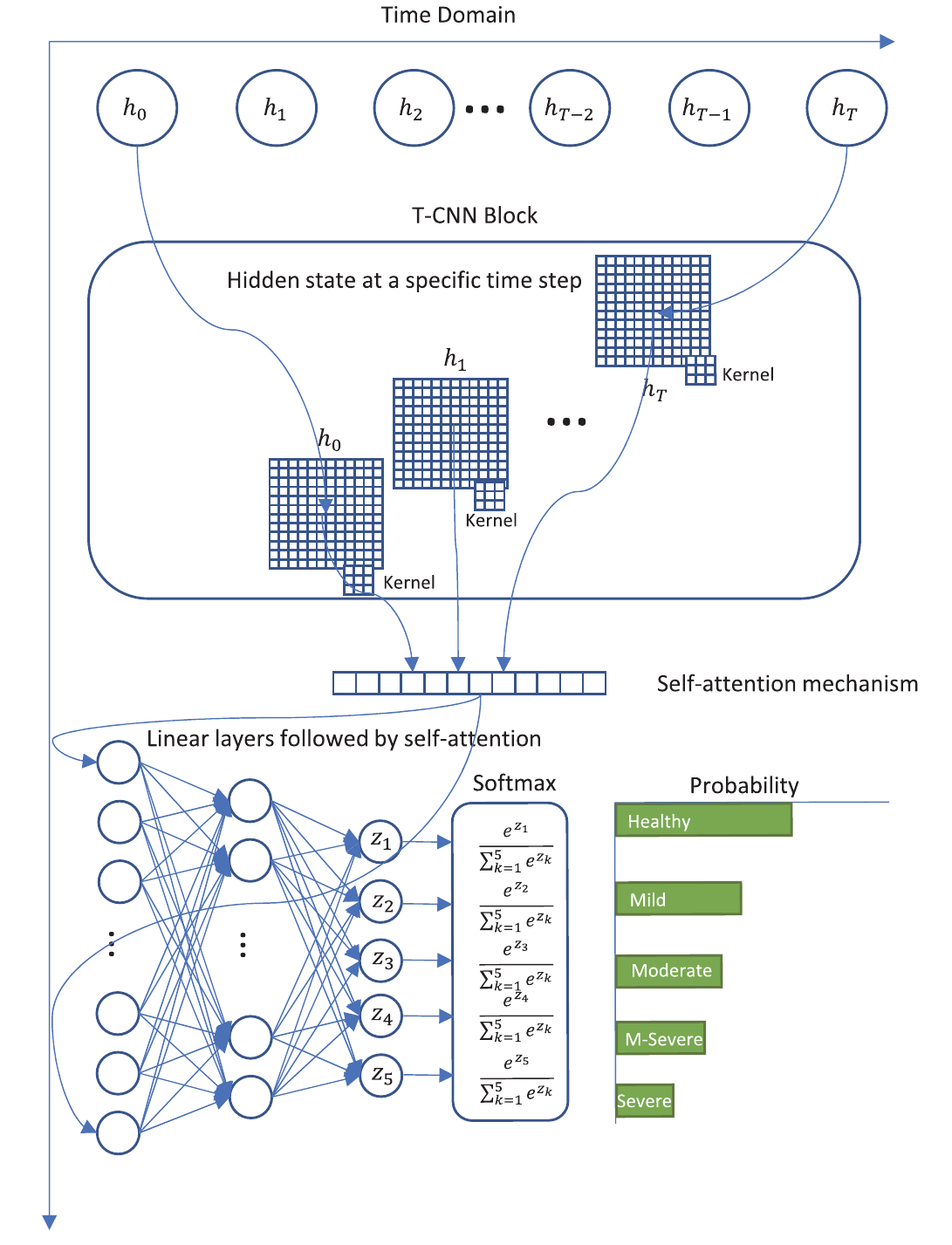}}
    \caption{The structure of the T-CNN model and the following linear neural network.}
    \label{fig:diagram_tcnn}
\end{figure}
Bidirectional LSTM is a variant of LSTM which consists of a forward layer on the original input sequence, and a backward layer on the reversed sequence. The Bi-LSTM outperforms the traditional LSTM because the forward and backward networks combine both forward and backward context information of the input sequence. Previous articles proposed to represent the input sequence by the last hidden state of the LSTM\cite{rodrigues2019multimodal,al2018detecting}. However, depression assessment is a complicated task, which heavily relies on the relationship between the audio features at different time steps, thus it is insufficient to use the last hidden state for classification, otherwise, it leads to the loss of temporal information. To solve this issue, we utilized the T-CNN to learn potential temporal and spatial information in the output of the Bi-LSTM. The structure of T-CNN is illustrated in Fig.\ref{fig:diagram_tcnn}. In general, simple CNN only supports the 2D or 3D spatial tensors as the input. However, the output shape of the LSTM is (\#samples, \#frames, \#LSTM neurons) given a unidirectional LSTM, and (\#samples, \#frames, 2*\#LSTM neurons) given a bidirectional LSTM. The T-CNN convolves the LSTM output vector along its 3rd axis and the shape of the convolution result is (\#samples, \#frames, \#output features, \#kernels). Therefore, we expand the shape of the LSTM output vector by inserting one new axis so that it can be processed by T-CNN. The T-CNN accepts a tensor with shape (\#samples, \#frames, 2*\#LSTM neurons, 1) as the input, which denotes a time series of LSTM hidden states. Our proposed T-CNN block consisted of three layers, first, time-distributed convolution layer, then time-distributed pooling layer to downsample the feature maps; and finally batch normalization layer. There were five T-CNN blocks in total in our proposed design, the output of the last T-CNN block contained "\#frame" samples, each sample is represented by 256 feature maps. Therefore, the last thing before the feature maps were fed into the following network was to downsample the output by the global average pooling layer, it slides along the time dimension of the feature vector and computes the mean value of each feature, which ensures that the relationship between each time step was taken into consideration. The output of the global average pooling layer was then fed into the following two linear layers. At last, the Softmax activation functions transformed neuron output into the probability of five severities. An Adam optimizer with a similar configuration in \ref{sec:lstm_text} was adopted for the training.
\subsection{Text Features and Models}
\label{sec:text}
The input layer of the text model took tokenized transcripts of each subject. Among all the 189 subjects in the dataset, text transcripts are in an average of 80 rows and a standard deviation of 14 rows. The interviews were in colloquial speech, thus the first step was to rephrase these colloquial descriptions to written languages, otherwise, colloquial terms all became out-of-vocabulary words, which were represented by the token [UNK], and greatly diminished model performance. 
Semantic information is highly essential in depression diagnosis because psychologists also formulate diagnosis by text produced by the patients during the interview. To acquire the text features, we firstly removed stop words in the patients' responses with Natural Language Toolkit (NLTK) and substituted some words and phrases such as "what's", "e-mail" with "what is" and "email", this eliminates different expressions of the same word \cite{bird2009natural}. Next, we lemmatized the remaining words in the sentences, the WordNet lemmatizer removes the inflectional endings and returns the base form of a word. Then the remaining texts were tokenized into word lists and were used to build a vocabulary with 7373 words. Each word in the vocabulary was assigned an index, the word list was then represented by these indices. After we acquired the word list, the main issue was that each word list was different in length, which made it more difficult to batch process text data if they were different in length. Therefore, the sliding window technique was applied to generate sequences in the same length, which was the same length as the sliding window. Each window consists of a constant number of words while 20\% words at the end were overlapping between two neighbouring time windows, which assigned higher weights to the words at the edge of the window so that the edge details were enhanced. The sliding window not only generated all training pairs but also performed data augmentation as well as directed the focus on a specific part of the sentence. Next, word sequences were encoded with the pre-trained 100D GloVe word embedding vector\cite{pennington2014glove}. The word embeddings were concatenated into a sentence embedding. For some short sentences, the size of the sliding window was greater than the length of the sentence, those short sentences were zero-padded to be the same length as the window. Therefore, the shape of the final input vector is (window size, 100). However, sentences shorter than 20\% of window size were discarded.
\subsubsection{Bi-LSTM text model}
Our proposed text model consists of a single-layer Bi-LSTM network and fully connected layers. The text feature sequences mentioned above comprise the index of words in the vocabulary. Text feature sequences were preprocessed to map each word to word embedding space with a non-trainable embedding layer before being fed into the model, and the shape of the embedding layer is (vocabulary size + 1, 100). Next, a batch normalization layer and then the Bi-LSTM layer further captured the semantic information underlying the input word sequences. To avoid overfitting, the dropout was applied to the recurrent input signal on the LSTM units and between fully-connected layers with the dropout rate of 0.2, and the shape of the Bi-LSTM output was (batch size, 200) at each time step. We adopted the attention mechanism to allow the model to adaptively select those depression-sensitive hidden states. The attention vector was then fed into two linear layers with 256 and 128 hidden units, respectively. Finally, the last linear layer with 5 hidden units determined the probability of the five severities. An Adam optimizer with a similar configuration in \ref{sec:lstm_text}  was adopted for the training. The cross-entropy loss calculated the distance between the output and the ground-truth label.

\subsection{Fused text-audio joint model}
\label{sec:fused}
Our final fused multimodality model was comprised of two subnetworks: text model and audio model, and followed by a shared late fusion neural network as Fig. \ref{fig:scheme_whole} shows. The late fusion neural network concatenated the outputs of the text and audio model to integrate text and audio features. For any subject, we extracted a high-level representation that included both semantic and prosodic features through the previous recurrent neural network and convolutional neural network. This high-level representation could be used in the following assessment of mental disorders. The output of our proposed model was a scoring matrix that denoted the likelihood of the depression severity. As the timesteps of the audio and text model were different, the late fusion network had to deal with input of different sizes. To solve this issue, we first attempted to adopt a max-pooling method to downsample the output from audio and text models so that they were in the same shape. Moreover, an attention mechanism was exploited, which provided us insights into the ratio of the contribution of each modality towards the final prediction.\par
Regarding fusion, we designed a set of models to integrate different modalities. Firstly, we fused the text models with different window sizes with the audio model with constant configuration. Our text model could be divided into two categories, one is the unidirectional LSTM text model, the other is the bidirectional LSTM. Our proposed audio and text model was previously described in Section \ref{sec:text} and Section \ref{sec:audio}, respectively. The only difference was that the output size of the audio and text model was 32 instead of 5 since they acted as feature extractors rather than classifiers. Global max pooling was adopted to align the extracted audio and text features. In order to integrate text and audio modalities, the output of the text and audio model was concatenated into a tensor and passed through a fully connected layer with 5 units. Secondly, the other fused model was set up using a similar configuration to the first one. The difference was that the attention mechanism played its role in aligning the features from different modalities. The third one was all the same as the previous two models, except it was created with an attention mechanism not only during the feature alignment but also in the fusion of the high-level representations. 

\section{EXPERIMENTS AND RESULTS}
In this section, the results of those models described in Section \ref{sec:dnm} are presented and discussed. We next assessed the effect of the hyperparameters for the proposed models. For the audio model, we compared the effect of architecture and timestep and investigated the potential long-term dependency of the audio features in severe patients. For the text model, we conducted experiments to investigate the effect of the hyperparameters such as the size of the window in preprocessing, the removal of stop words. Regarding the audio-text fused model, we mainly focused on the impact of fusion methods on the model performance. All the experiments were conducted on one RTX 2080Ti 11GB GPU. The size of multimodality models was limited mainly by the amount of memory available on our GPU and the amount of time for training we can tolerate. Our single-modality model usually took between 3 to 5 hours to train, but the training of our proposed multimodality model always took around 20 hours. The results of our experiment provided an insight that our models could be improved by faster GPUs and larger datasets. The detailed results are discussed in the following parts.
\subsection{The Statistics of Audio and Text Features}
The pause time between responses is also longer than usual in the depressive population \cite{kraepelin1921manic}. To verify whether the DAIC-WOZ dataset follows a similar pattern, we calculated the statistics of the raw interview recordings and the transcripts. The subjects were divided into two groups by PHQ-8 scale, the subjects were considered as normal or mild (control group) if their PHQ-8 is less or equal to 10, otherwise, they are considered as moderate or severe (experiment group). This threshold is given by a previous study on the efficacy of PHQ-8 on the diagnosis of major depressive disorder. It was reported that given the cutoff score of 10, the PHQ-8 exhibited a sensitivity of 58.3\%, specificity of 83.1\%\cite{shin2019comparison}. The two-sided T-test was applied to test if there was a significant difference in the audio duration between the control and experiment groups. The statistics of the two groups are listed in the Table. \ref{table:result_t_test}. The histograms of the audio duration and sentence length of the control and experiment groups are illustrated in Fig. \ref{figure:audio_sentence_length}. The response duration of the control and experiment groups is on an average of 951.3711 $\pm$ 266.6010 and 997.8773 $\pm$ 290.1901 seconds, respectively. The two-tailed p-value is 0.0952. The sentence length of the control and experiment groups is on average of 8.7854 $\pm$ 8.9475  and 7.3717 $\pm$ 7.2975 in the number of words, respectively. The two-sided T-test was applied to test if there was a significant difference between the sentence length in the control and experiment groups. The two-tailed p-value is $3.2397 \times 10^{-14}$. The above results indicate no significant difference in the audio duration of the control and experiment groups. However, the sentence lengths for the control and experiment groups are significantly different. More responses in the experiment group consisted of less than 5 words. As the audio durations between the control and experiment groups have identical average values, we can conclude that there are more pauses in the conversations of the experiment group. This result is identical to other researchers’ conclusions. Therefore, our dataset and criterion for depression are reasonable.
\subsection{Results of the Audio Modality}

\begin{table}[htb]
\centering
\caption{Results of the Baseline Audio Models}
\label{result:baseline_models}
\resizebox{\columnwidth}{!}{
\begin{tabular}{ccccc}
\toprule
          & \multicolumn{4}{c}{Test Baseline}                                                                       \\
Models    &  \multicolumn{2}{c}{Random Forest} & Madhavi et al.\cite{madhavi2020deep}. & Yang et al.\cite{yang2017hybrid} \\
\midrule
          & Mean            & St. dev         & Mean            & St. dev                   \\
Accuracy            & 0.3192          & 0.0085          & 0.7500         & 0.8273          \\
Precision           & 0.3206          & 0.0064          & 0.7200         & 0.7930   \\
Recall              & 0.3184          & 0.0040          & 0.7500         & 1.0000   \\
F1 Score            & 0.3168          & 0.0076          & 0.7300         & 0.8850  \\
\bottomrule
\end{tabular}
}
\end{table}

\begin{figure*}[htb]
\centering
\begin{subfigure}[t]{0.33\textwidth}
\centering
  \includegraphics[width=\textwidth]{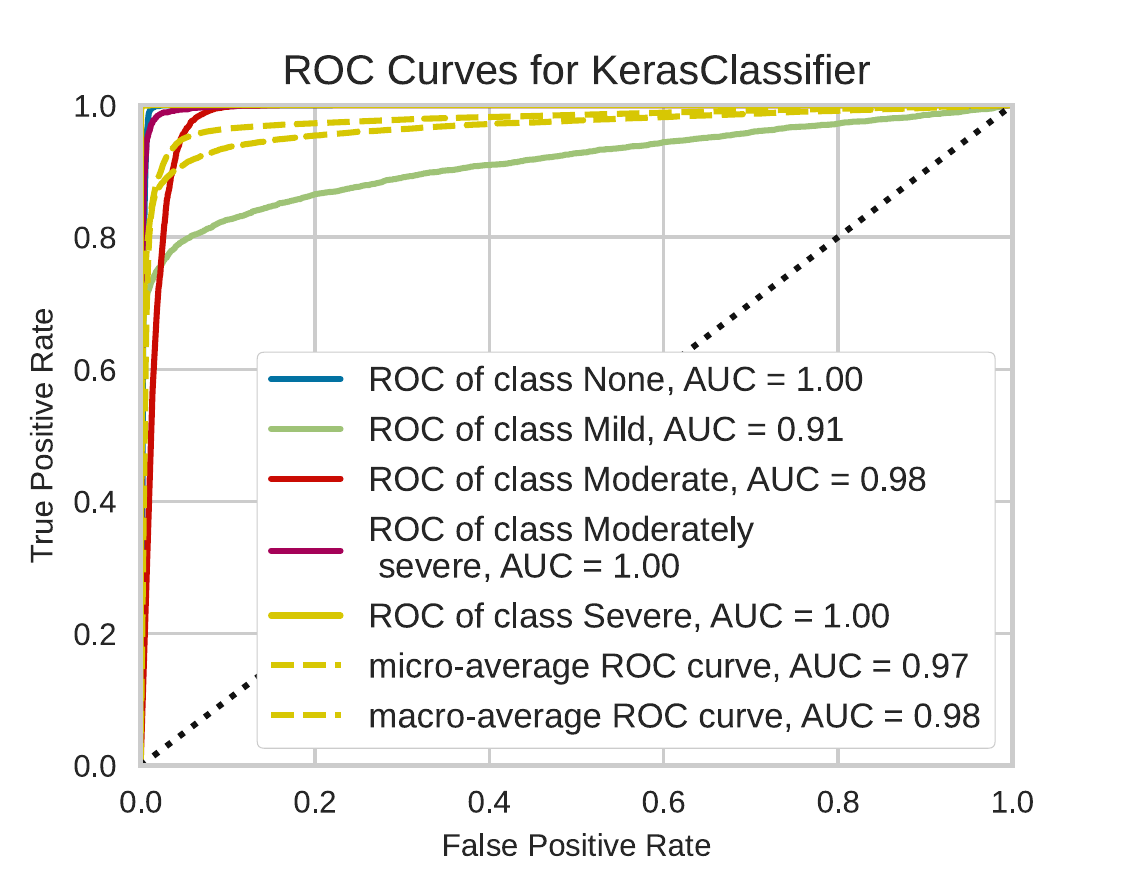}
\caption{The ROC of 16-timestep model on DAIC-WOZ}
\label{fig:audio_roc_16}
\end{subfigure}
\begin{subfigure}[t]{0.33\textwidth}
\centering
\includegraphics[width=\textwidth]{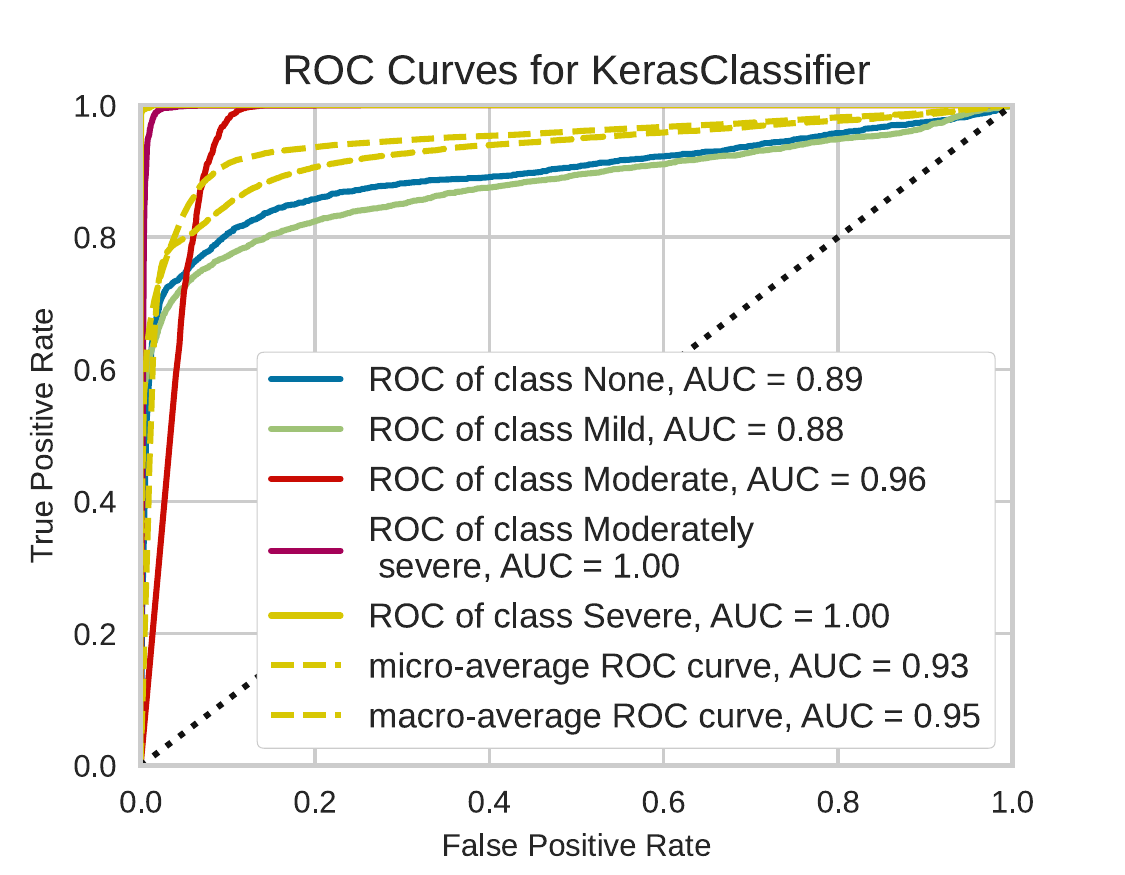}
\caption{The ROC of 32-timestep model on DAIC-WOZ}
\label{fig:audio_roc_32}
\end{subfigure}
\begin{subfigure}[t]{0.33\textwidth}
\centering
\includegraphics[width=\textwidth]{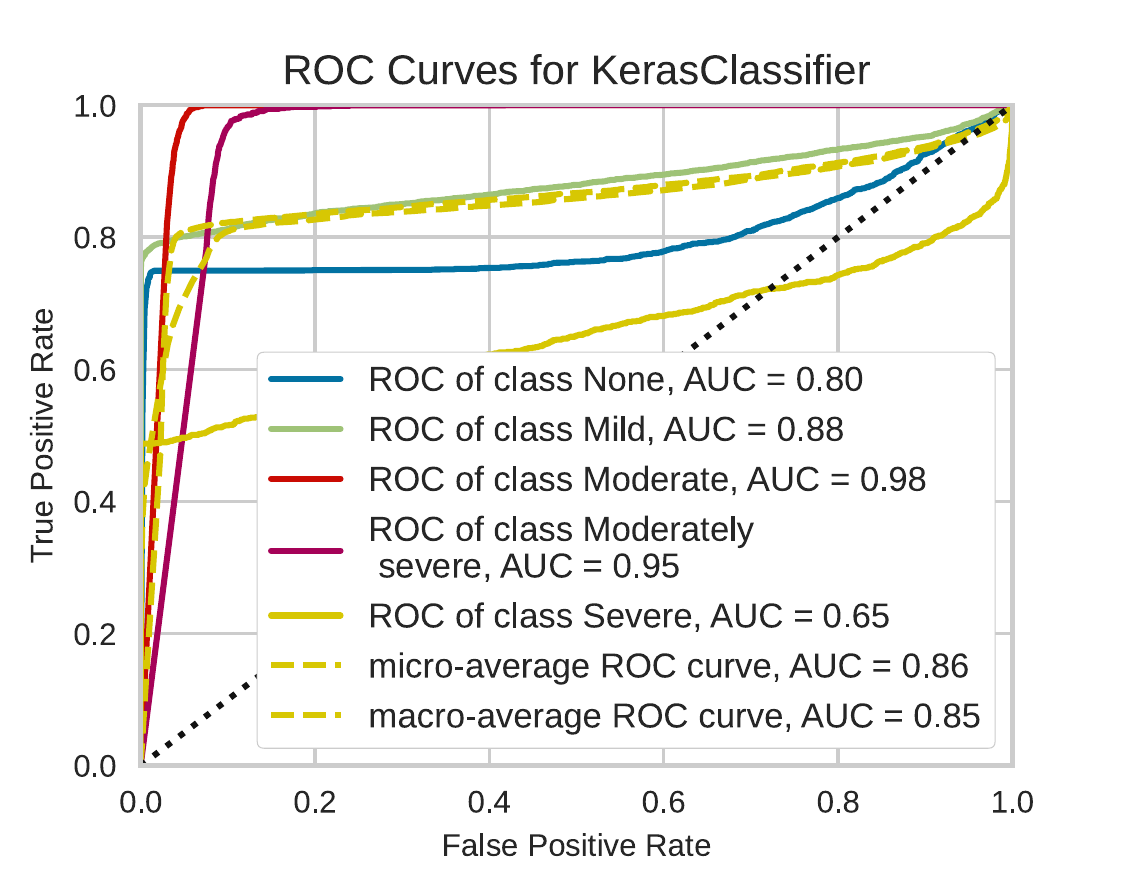}
\caption{The ROC of 64-timestep model on DAIC-WOZ}
\label{fig:audio_roc_64}
\end{subfigure}
\caption[]
        {\small The ROC of three different model configurations. (\ref{fig:audio_roc_16}) The Bi-LSTM followed by T-CNN given the time step = 16. Micro-Average AUC: 0.97. The AUC of “Severe” is smaller than any other class, this indicates the detection of severe depression is more challenging than other depression levels. (\ref{fig:audio_roc_32}) The Bi-LSTM followed by T-CNN given the time step = 32. Micro-Average AUC is 0.93. The micro-average AUC is smaller compared with that when the time step = 16. The longer sequence does not mean a better result because the noise introduced by the longer sequence can mislead the model. (\ref{fig:audio_roc_64}) The Bi-LSTM followed by T-CNN given the time step = 32. Micro-Average AUC is 0.86, which is in line with our expectation that a longer input sequence makes it more challenging to predict the severity.}\centering
\label{fig:roc_audio_daic}
\end{figure*}

\begin{figure*}[bp]
\centering
 
\begin{subfigure}[t]{0.33\textwidth}
\centering
\scalebox{0.8}{\includegraphics[width=\textwidth]{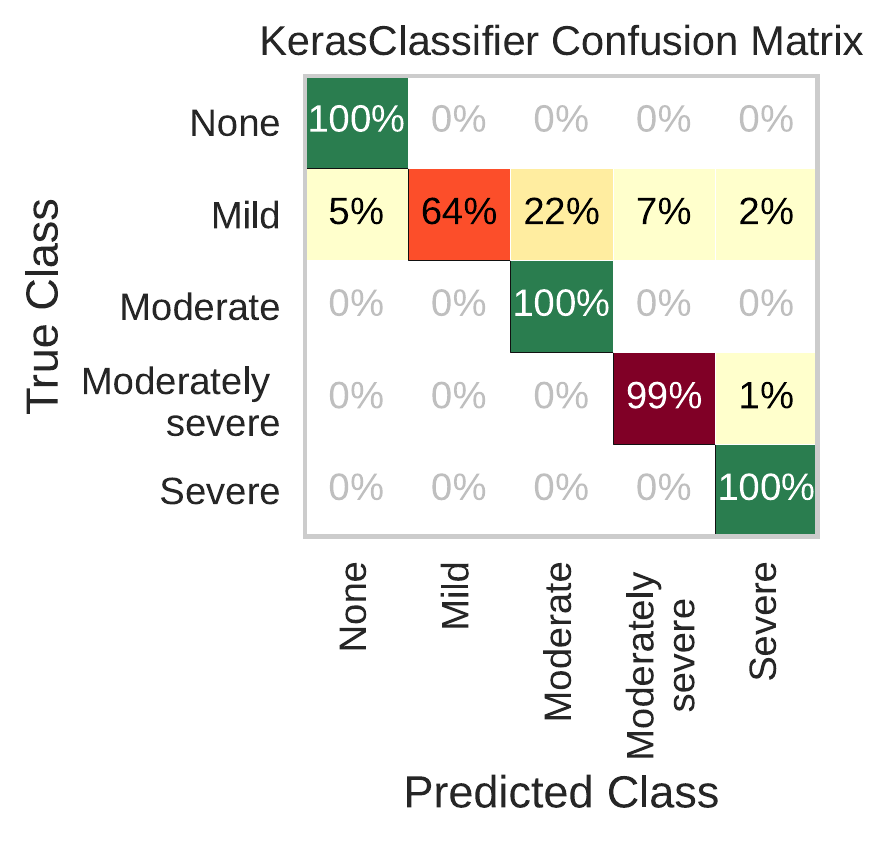}}
\caption{Confusion matrix of 16-timestep model on DAIC-WOZ}
\label{fig:daic_woz_16_cm}
\end{subfigure}
\begin{subfigure}[t]{0.33\textwidth}
\centering
\scalebox{0.8}{\includegraphics[width=\textwidth]{fig7a.pdf}}
\caption{Confusion matrix of 32-timestep model on DAIC-WOZ}
\label{fig:daic_woz_32_cm}
\end{subfigure}
\begin{subfigure}[t]{0.33\textwidth}
\centering
\scalebox{0.8}{\includegraphics[width=\textwidth]{fig7a.pdf}}
\caption{Confusion matrix of 64-timestep model on DAIC-WOZ}
\label{fig:daic_woz_64_cm}
\end{subfigure}
\caption{Results of the proposed audio model on the DAIC-WOZ dataset}
\label{fig:audio_res_daic_woz}
\centering
\end{figure*}

As for the audio models, evaluation metrics accuracy, recall, precision, and f1 score used to evaluate models with different configurations are shown in Tables \ref{result:baseline_models} and \ref{result:proposed_models}. The test set for evaluation is balanced by oversampling the minority class.
Random forest was used as the baseline in evaluating the audio modality sequence-level prediction. Audio feature sequences for training and evaluating are non-stationarity series, which are difficult to model and forecast. They were pre-processed by differencing to be made stationary. Differencing is the change from one audio feature sampling time to the next. The random forest model we used in this manuscript is an ensemble approach that fits a set of decision trees on different sub-sample of the dataset, and averaging the output of each decision tree to improve the prediction accuracy, as well as prevent the model from overfitting. In our article, 100 decision trees were trained on various sub-sample of the training set to construct the random forest model. Another baseline method, Madhavi et al. proposed a CNN consisting of 2 convolutional layers and two successive linear layers to extract high-level features from the frequency spectrogram of interview recordings. The output of CNN is fed into the following neural networks to predict an individual’s depression level. They also evaluated their models on the DAIC-WOZ dataset. Moreover, Yang et al. proposed a similar but more complex model, they also adopted the combination of convolution neural networks and deep neural networks (i.e. multi-layer perceptron model). Each subject was labelled by their depression-related symptoms, such as prior depression diagnosis, sleep disorder, present or not. Their proposed CNN consists of three convolution layers and the intermediate output of CNN is fed into the deep neural network to predict the presence of depression symptoms. These symptom labels are fed into another deep neural network for predicting depression severity. Their results on the DAIC-WOZ dataset are summarized in our comparative studies.
For the LSTM with the fully connected layers model, it outperformed the baseline model machine learning model (i.e. decision tree) by 24\% in terms of accuracy. In contrast, the  Bi-LSTM with the fully connected layers model outperformed by 54\% in terms of accuracy. For our proposed Bi-LSTM combined with the T-CNN model, we achieved 16\% improvements over the best baseline model in terms of accuracy. From Tables \ref{result:baseline_models} and \ref{result:proposed_models}, it can be concluded that the LSTM performed better on the depression level classification compared with the baseline machine learning models, such as the naïve Bayes model. Moreover, we observed that the network followed by the LSTM layer is critical for good performance. If the other configurations were fixed, Bi-LSTM with T-CNN outperformed other methods because the T-CNN learned more temporal and spatial information than others by capturing the correlation within all hidden states of the LSTM. We also investigated the influence of the value of the time step and concluded that our model performed best when the timestep was 16. Fig. \ref{fig:audio_roc_16} shows the receiver operating characteristic (ROC) curve when timestep=16. The micro-average AUC for our proposed model is 0.98, and the AUC for “mild” is smaller than any other, which indicates it is more challenging for the model to distinguish mild depression from the other levels correctly. This is likely to be attributed to the fact that mild patients behave very similarly to healthy people during the interview. 
Fig. \ref{fig:audio_roc_32} is the ROC when the time step is 32. The micro-average AUC for this model is 0.93. The performance of the model with 32-timesteps was worse than that of the model with 16-timesteps. This is likely due to the negative correlation between the signal-noise ratio of the input sequence and the length of the sequence. A longer input sequence contains more information to assess the emotional state, but as the sequence length grows, the increasing noise cannot be ignored and the bias of the model rises due to the noise. Another factor is the limitation of the memorization capability of LSTM. The longer the input sequence is, the more difficult it is for LSTM to memorize earlier information when processing the end of the sequence because the depth of the LSTM network is proportional to the timestep. Given a long sequence, the information cannot smoothly flow through the network, which results in diminished performance. The confusion matrix of the 32-time step model is illustrated in Fig. \ref{fig:daic_woz_32_cm}, which shows the performance of the model on the test partition of the DAIC-WOZ dataset. Comparing the models with different time steps, Fig. \ref{fig:daic_woz_16_cm} shows the confusion matrix of the model with 16 timesteps, while Fig. \ref{fig:daic_woz_64_cm} shows the confusion matrices of the model with 64 timesteps. Different timestep means the different sizes of the test set. To eliminate the influence of the size of the test set, we normalized the confusion matrix along each row. In terms of the normalized confusion matrix, the model with 16 timesteps performed the best, but from the entries on the second row of Fig. \ref{fig:daic_woz_64_cm}, the model with 64 timesteps was less likely to classify the mild patients incorrectly. The contribution of the model with a longer time step in the depression prediction should be further investigated to find the cut-off value of the time step that optimizes the trade-off between the computation cost (larger time step means more computation) and the misdiagnosed rate.

\subsection{Results of the Text Modality }

\begin{table*}[htb]
\centering
\caption{A Comparative Study of Different Proposed Audio Models}
\resizebox{\textwidth}{!}{%
\begin{threeparttable}
\begin{tabular}{@{}llccc@{}}
\toprule	
Models & \multicolumn{1}{c}{Experimental Settings} & \multicolumn{1}{c}{Accuracy} & \multicolumn{1}{c}{F1} & \multicolumn{1}{r}{p-value}  \\ \midrule
LSTM + FC & TS=16, HU=73, LHU=(128,64,5), Adam & $0.5674\pm0.0034$ & $0.5650\pm0.0042$ & $\textless.01$ \\
Bi-LSTM + FC & TS=16, HU=73, LHU=(128,64,5), Adam & $0.8717\pm0.0013$ & $0.8818\pm0.0013$ & $\textless.01$ \\
LSTM + T-CNN & TS=16, HU=73, \#TCNNB=5, \#KRNL=(64,64,64,128,256), KS=(3,3,3,3,9), Adam & $0.8698\pm0.0897$ & $0.8609\pm0.0988$ & $\textless.01$ \\
\textbf{Bi-LSTM + T-CNN} & \textbf{TS=16, HU=73, \#TCNNB=5, \#KRNL=(64,64,64,128,256), KS=(3,3,3,3,9), Adam} & $\mathbf{0.9871\pm0.0009}$ & $\mathbf{0.9870\pm0.0009}$ & $\mathbf{\textless.01}$ \\ \bottomrule
\end{tabular}%
\footnotesize
\begin{tablenotes}
\item[] \#TCNNB: Number of T-CNN blocks \quad \#KRNL: Number of conv kernels in each T-CNN block \quad \#KS: Kernel size
\end{tablenotes}
\end{threeparttable}
}
\label{result:proposed_models}
\end{table*}

\subsubsection{The effect of stop words and bidirectional layer}
In this experiment, we used NLTK to remove the stop words in English transcripts.
Apart from the stop words, the other factor is the choice between LSTM and Bi-LSTM models. Compared with the unidirectional LSTM model, the bidirectional model converges faster, and the validation accuracy is higher. The following experiment demonstrates several advantages of the Bi-LSTM model over the traditional LSTM model on the depression level classification task. Four models were trained with the different configurations presented in Table. \ref{result:text_model}. The test set for evaluation is balanced by oversampling the minority class.

\begin{table*}[htb]
\centering
\caption{A Comparative Study of the Proposed Text Models}
\resizebox{\textwidth}{!}{%
\begin{threeparttable}
\begin{tabular}{@{}llccc@{}}
\toprule
Models & \multicolumn{1}{c}{Experimental Settings} & \multicolumn{1}{l}{Accuracy} & F1 & \multicolumn{1}{r}{Micro-average AUC} \\ \midrule
LSTM + FC & TS=64, HU=100, LHU=(256,128,5), Adam, Stopwords & 0.9091 & 0.9094 & 0.9738 \\
LSTM + FC & TS=64, HU=100, LHU=(256,128,5), Adam, No stopwords & 0.9792 & 0.9754 & 0.9897 \\
Bi-LSTM + FC & TS=64, HU=100, LHU=(256,128,5), Adam, Stopwords & 0.9617 & 0.9610 & 0.9908 \\
\textbf{Bi-LSTM + FC} & \textbf{TS=64, HU=100, LHU=(256,128,5), Adam, No stopwords} & \textbf{0.9685} & \textbf{0.9709} & \textbf{0.9925} \\ \bottomrule
\end{tabular}%
\footnotesize
\begin{tablenotes}
\item[] TS: Timestep;\quad HU: \#Hidden units in LSTM;\quad LHU: \#Hidden units in linear layers 
\end{tablenotes}
\end{threeparttable}
}
\label{result:text_model}
\end{table*}
From Table. \ref{result:text_model}, we concluded that if the type of the LSTM was fixed (i.e., the two text models both consist of LSTM or Bi-LSTM network), the performance of the model without stop words was better. If the stop words were kept, the Bi-LSTM model still outperformed the traditional one. This result was in line with our expectation that Bi-LSTM was better in text classification because it learned more contextual information with the combination of the forward and backward networks.
\subsubsection{The effect of window size}
Window size is another factor that influences the performance of the model. Intuitively, the longer the window, the more information it contains about the mental state of the subjects, which means our model can assess their emotions more accurately. However, if the window is too long, while making an inference, the impact of the noise cannot be ignored, which leads to significant performance degradation. Moreover, the memorization capability of LSTM is limited, which means the longer the sequence is, the more challenging for the LSTM to memorize and extract useful information. To demonstrate the relationship between the performance and the window size, we conducted experiments by changing the window size. As shown in Table \ref{result:text_model_win_size}, when the window size 
started to increase, the metrics increased firstly but began to decrease after the window size is greater than 64. This was in line with our expectation, the classifier gained a lot of information due to a larger window but started to degrade as the result of the noise in the large window and the reduced performance of LSTM. We concluded that the window size should be appropriately set to train the model with the best performance, in our experiment, the best window size is 64.
\begin{table}[htb]
\centering
\caption{A Comparative Study of the Text Model with Different Window Size}
\label{result:text_model_win_size}
\scalebox{1.0}{\begin{tabular}{lllll}
\toprule
\makecell[c]{Window Size} & \makecell[c]{Accuracy} & \makecell[c]{Precision} & \makecell[c]{Recall} & \makecell[c]{F1 Score}\\
\midrule
\makecell[c]{16} & \makecell[c]{0.8254} & \makecell[c]{0.8318} & \makecell[c]{0.8340} & \makecell[c]{0.8141}    \\
\makecell[c]{32} & \makecell[c]{0.8256} & \makecell[c]{0.8371} & \makecell[c]{0.8465} & \makecell[c]{0.8260}   \\
\makecell[c]{\textbf{64}} & \makecell[c]{\textbf{0.8778}} & \makecell[c]{\textbf{0.8779}} & \makecell[c]{\textbf{0.8782}} & \makecell[c]{\textbf{0.8705}}   \\
\makecell[c]{128} & \makecell[c]{0.8409} & \makecell[c]{0.8599} & \makecell[c]{0.8430} & \makecell[c]{0.8304}  \\
\bottomrule
\end{tabular}}
\end{table}
\subsection{Results of the fused model}
\begin{table*}[htb]
\caption{A Comparative Study of Our Proposed Patient-Level Methods and the State of The Art}
\label{table:final_result}
\centering
\scalebox{1.0}{
\resizebox{\textwidth}{!}{%
\begin{tabular}{@{}llcccc@{}}
\toprule
Model & Experimental Settings & Accuracy & F1 & Sensitivity & Specificity \\ \midrule
\multirow{3}{*}{UniLSTM as encoder} & WIN=16, Stride=64 & 0.8604 & 0.8579 & 0.9844 & 0.8182 \\
 & WIN=32, Stride=64 & 0.9209 & 0.9188 & 0.9647 & 0.9777 \\
 & WIN=64, Stride=64 & 0.8674 & 0.8682 & 0.9705 & 0.9888 \\ \midrule
\multirow{3}{*}{BiLSTM as encoder} & WIN=16, Stride=64 & 0.9488 & 0.9500 & 0.9735 & 0.9444 \\
 & WIN=32, Stride=64 & 0.9186 & 0.9191 & 0.9852 & 0.9700 \\
 & WIN=64, Stride=64 & 0.8535 & 0.8546 & 0.9647 & 0.8778 \\ \midrule
\multirow{3}{*}{BiLSTM as encoder} & WIN=16, Stride=64, attention & 0.8419 & 0.8427 & 0.9735 & 0.8222 \\
 & \textbf{WIN=32, Stride=64, attention} & \textbf{0.9581} & \textbf{0.9580} & \textbf{0.9824} & \textbf{1.0000} \\
 & WIN=64, Stride=64, attention & 0.9093 & 0.9086 & 0.9706 & 0.9889 \\ \midrule
\multirow{3}{*}{UniLSTM as encoder} & WIN=16, Stride=64, attention(aligning\&fusion) & 0.8977 & 0.8973 & 0.9559 & 0.9889 \\
 & WIN=32, Stride=64, attention(aligning\&fusion) & 0.9326 & 0.9315 & 0.9735 & 0.9889 \\
 & WIN=64, Stride=64, attention(aligning\&fusion) & 0.8581 & 0.8615 & 0.9412 & 0.8889 \\ \midrule
\multirow{3}{*}{BiLSTM as encoder} & WIN=16, Stride=64, attention(aligning\&fusion) & 0.8491 & 0.8439 & 0.9353 & 0.9000 \\
 & WIN=32, Stride=64, attention(aligning\&fusion) & 0.9047 & 0.9103 & 0.9941 & 0.9000 \\
 & WIN=64, Stride=64, attention(aligning\&fusion) & 0.6279 & 0.6560 & 0.7500 & 1.0000 \\ \midrule
 \multirow{4}{*}{Unimodality text model} & WIN=16 & * & BLSTM: 0.7929\quad ULSTM:0.8096 & * & * \\ 
 & WIN=32 & * & BLSTM: 0.7964\quad ULSTM:0.7619 & * & * \\ 
  & \textbf{WIN=64} & * & \textbf{BLSTM: 0.9245\quad ULSTM:0.9058} & * & * \\ 
   & WIN=128 & * & BLSTM: 0.8266\quad ULSTM:0.7148 & * & * \\ 
 \midrule
 \multirow{4}{*}{Unimodality audio model} & BLSTM + FC & * & 0.8819 & * & * \\
 & ULSTM + FC & * & 0.7604 & * & * \\
 & \textbf{BLSTM + TCNN} & * & \textbf{0.9074} & * & * \\
 & ULSTM + TCNN & * & 0.8443 & * & * \\ \midrule
 Srimadhur et al.\cite{srimadhur2020end} & End to end convolutional neural network & 0.7464 & 0.7750 & 0.74 & 0.8 \\
Alhanai et al.\cite{al2018detecting} & Combination of LSTM and CNN & * & 0.77 & 0.83 & * \\
 Niu et al.\cite{niu2021hcag} & Hierarchical context-aware graph attention model & * & 0.92 & 0.92 & * \\ \bottomrule
\end{tabular}%
}
}

\end{table*}
In this experiment, the audio and text models were jointly optimized so that we could verify whether our methods were still effective under multimodality configuration. We proposed three varieties of fusion models and merged these segment-wise predictions through major voting to obtain the patient-level prediction. The configuration details of those fused models were described in Section \ref{sec:fused}. The metrics of each fusion model on the test partition were covered in Table. \ref{table:final_result}. When experimenting with models made up of unidirectional LSTM, without an attention mechanism, the model with a window size of 32 performed better than others when classifying for a multi-class outcome in terms of the accuracy on the test set($accuracy=$ 0.9209). Theoretically, the models with Bi-LSTM should be better than a uni-LSTM one, however, with all other configurations fixed, except the Bi-LSTM model with a window size of 16, other Bi-LSTM models did not show significant improvement over the uni-LSTM one. Nevertheless, once the attention mechanism was introduced, the performance was boosted and the $F_1$ increased compared to the model without an attention mechanism, except the Bi-LSTM model with an attention mechanism and window size of 16. As we reported in the methodology section, the attention mechanism could be introduced during the multimodal feature aligning phase as well as the multimodality fusion phase. The attention mechanism during the fusion process weighed each modality and made it possible for the model to determine the contribution of each modality. From Table. \ref{table:final_result}, we concluded that the highest sensitivity of 0.9941 was achieved by the model comprised of Bi-LSTM and two attention layers, with a window size of 32. Given that we expected to train an early-stage depression screening tool, we preferred higher sensitivity so that we would not miss those potential depression patients. The model with two attention layers led to results that outperformed the state of the art, Niu et al., by 8\% in terms of sensitivity. In comparison with Alhanai et al., who adopted a similar method made up of CNN and LSTM, our proposed method was better by 17\% in terms of sensitivity. This is not conclusive since the dataset for evaluation in their article was slightly different from ours. By conducting a student t-test between the $F_1$ of the best patient-level audio model with the result of p-value = 0.0099 (\textless0.01), as well as patient-level text model with the result of p-value = 0.0246 (\textless0.05), we could conclude that multimodality models statistically significant outperformed single modality models.

\section{CONCLUSION}
In this paper, a multimodality approach for automated depression detection was presented. Firstly, we performed the statistical test to investigate the difference between the audio and text features of severe and healthy subjects. We proved the pattern of severe depression patients was different from that of the healthy. Therefore, the audio feature sequence carried information that could be used to predict depression severity.
Secondly, models that considered audio and text features individually were trained and evaluated at the patient-independent level. These unimodality models then acted as feature extractors and output features were combined by audio-text fused model. For the audio modality, at the patient-independent level, the model comprised of single-layer Bi-LSTM and five stacked T-CNN blocks achieved the best sequence level $F_1$ score of 0.9870 and patient-level $F_1$ score of 0.9074 with the test set. This result indicates that the Bi-LSTM provides a more reliable representation, from which the automated depression detection model could benefit. Additionally, we evaluated the patient-independent audio models with different timesteps with the Area Under Curve (AUC) metric. We concluded that the 16-timestep model performed best and the micro-average AUC was higher than any other model. However, the 64-timestep model showed its strength in detecting the audio feature sequence from the mild patient, which met our expectation that the model should be able to distinguish mild patients so that clinical interference can be conducted in the early stage. Overall, the 16-timestep model outperformed the 32-timestep and 64-timestep models, which could be attributed to the relatively low signal-noise ratio of the shorter input sequences and the memorization limit of the LSTM. The new understanding assisted in our model selection and hyper-parameter configuration when we deployed this method in clinical settings. These findings provided the following insight for future research, our proposed unimodality model was patient-independent, and the prediction was based on a period of audio/text features. Therefore, compared with other models, our proposed model did not have limitations to the length of interview audio or transcript, which made it possible for people to monitor their mental state in daily use. 

Moreover, for the text modality, the model consisted of Bi-LSTM and three fully connected layers achieved the best sequence level $F_1$ score of 0.9709 and patient-level $F_1$ score of 0.9245 on the test set. We conducted experiments to investigate the influence of the text model hyper-parameters, such as window size and stop words. We found the best window size is 64. In our experiment, we investigated the effect of stop words, the result indicated the text model performs better if the stop words were removed in advance.
Currently, our patient-level prediction was carried out by a major voting algorithm, which yielded a patient-level depression prediction model with satisfying performance. Our proposed multimodal method achieved the highest $F_1$ of 0.9580 on the patient-level depression detection task, which showed a significant improvement over the previous state-of-the-art. In the future, a study on how to represent the audio/text features during the whole interview should be carried out so that the model could make patient-level predictions based on a digest of text and audio features.

\ifCLASSOPTIONcompsoc
  \section*{Acknowledgments}
\else
  \section*{Acknowledgment}
\fi

We would like to acknowledge the funding support from the MITACS grant, Canada. This research is also supported by the China Scholarship Council (CSC) No. 202000810031. The authors would also like to thank Jo-Sheen Yen for proofreading the manuscript.

\ifCLASSOPTIONcaptionsoff
  \newpage
\fi



\bibliographystyle{IEEEtran}
\bibliography{IEEEabrv,./ref.bib}

% or


% \appendices
\begin{IEEEbiography}[{\includegraphics[width=1in,height=1.25in,clip,keepaspectratio]{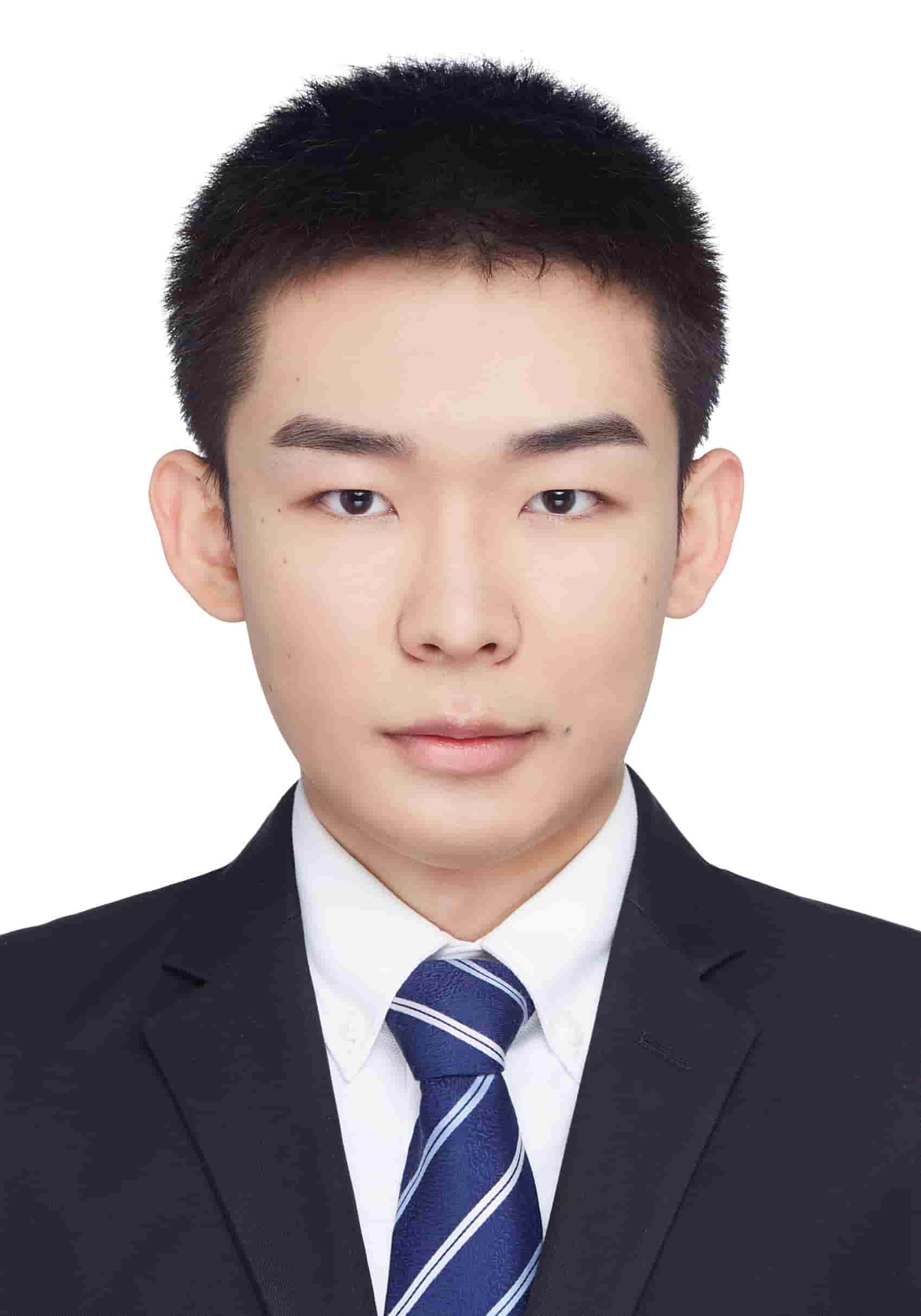}}]{Kaining Mao}
Kaining is a Ph.D. student in the Electrical Computer Engineering Department at the
University of Alberta. His work focuses specifically on Multimodal Machine Learning
(MMML) and its impact on Automated Depression Diagnosis (ADD).
\end{IEEEbiography}

\begin{IEEEbiography}[{\includegraphics[width=1in,height=1.25in,clip,keepaspectratio]{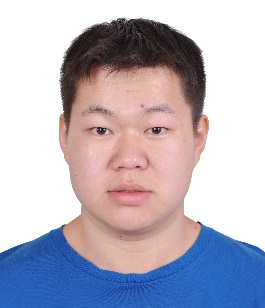}}]{Wei Zhang}
Wei Zhang is a Ph.D.
student in Electrical
Computer Engineering.
He is working on
artificial
intelligence field to
detect tuberculosis and
depression in different
projects.
\end{IEEEbiography}

\begin{IEEEbiography}[{\includegraphics[width=1in,height=1.25in,clip,keepaspectratio]{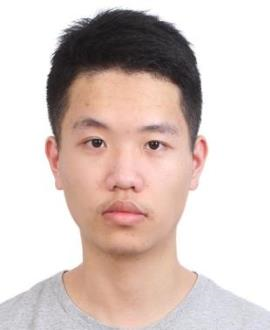}}]{Ang Li}
Ang Li is a BSc student
in Computer
Engineering at the
University of Alberta.
He is currently
conducting the Dean
Research Awards
Program focuses on
the application of AI in
depression recognition.
His research interests
include machine
learning and
embedded system
design.
\end{IEEEbiography}
\begin{IEEEbiography}[{\includegraphics[width=1in,height=1.25in,clip,keepaspectratio]{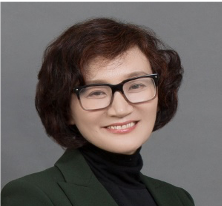}}]{Deborah Baofeng Wang}
Dr. Deborah Baofeng Wang is responsible for the management of several international R\&D projects at Wenzhou Kangning Hospital, focusing on the application of AI in mental health. With a Ph.D. in educational psychology and an MBA/management information system concentration, Dr. Wang has extensive experiences working as a Senior Research Analyst in the fields of mental health, public health, and education handling longitudinal and cross-sectional data at national (US) and international levels.
\end{IEEEbiography}

\begin{IEEEbiography}[{\includegraphics[width=1in,height=1.25in,clip,keepaspectratio]{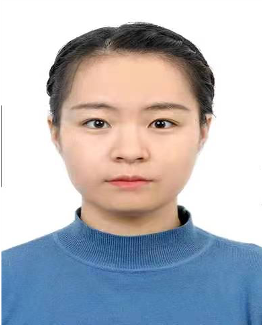}}]{Rongqi Jiao}
Rongqi Jiao is currently a Master's student at Wenzhou Medical University. With a Bachelor's degree in Medicine and the opportunity to learn as a psychiatrist at Wenzhou Kangning Hospital, Ms. Jiao has systematic academic training in medicine and hands-on research and clinical experiences in the area of mental health.
\end{IEEEbiography}

\begin{IEEEbiography}[{\includegraphics[width=1in,height=1.25in,clip,keepaspectratio]{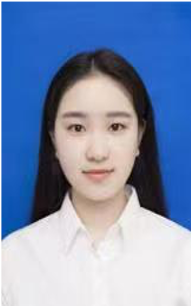}}]{Yanhui Zhu}
Yanhui Zhu is a Master's student majoring in psychiatry and mental health in the School of Mental Health, Wenzhou Medical University. The primary focus of her research in the Master's program has been on microexpression and mental illness. She is currently working in the Second Affiliated Hospital of Wenzhou Medical University, as a clinical practitioner in mental health.
\end{IEEEbiography}

\begin{IEEEbiography}[{\includegraphics[width=1in,height=1.25in,clip,keepaspectratio]{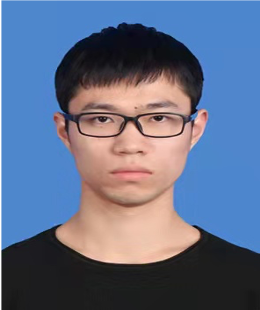}}]{Bin Wu}
Bin Wu is currently a Master's student at Wenzhou Medical University. He holds a Bachelor’s degree in Medicine and is now also working as a psychiatrist at Wenzhou Kangning Hospital. He has received systematic academic training in the area of mental health and has solid hands-on experience in clinical practice of mental health, particularly the application of AI in mental health.
\end{IEEEbiography}

\begin{IEEEbiography}[{\includegraphics[width=1in,height=1.25in,clip,keepaspectratio]{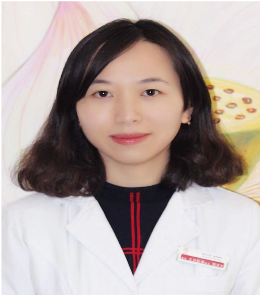}}]{Tiansheng Zheng}
Dr. Tiansheng Zheng has a Master's degree in psychiatry and has been accredited as an Associate Chief Physician in psychiatry. She currently serves as the Deputy Director of Depression Center and Sleep Center of the Affiliated Kangning Hospital of Wenzhou Medical University; Her research interests include cognitive function in patients with depression and insomnia disorders.
\end{IEEEbiography}

\begin{IEEEbiography}[{\includegraphics[width=1in,height=1.25in,clip,keepaspectratio]{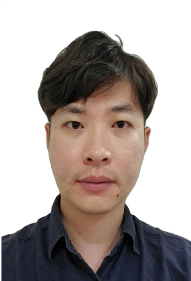}}]{Lei Qian}
Mr. Lei Qian holds a Master's degree in clinical psychology and is accredited as an Intermediate Psychotherapist. He also serves as the Deputy Director of Higher Education Collaborations with the Affiliated Kangning Hospital of Wenzhou Medical University. His research interests include data analysis and mental health.
\end{IEEEbiography}

\begin{IEEEbiography}[{\includegraphics[width=1in,height=1.25in,clip,keepaspectratio]{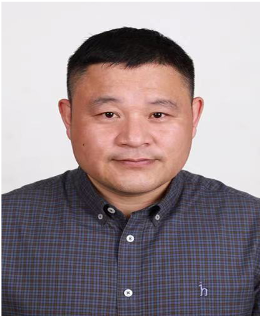}}]{Wei Lyu}
Accredited as a Chief Physician in psychiatry and Intermediate Psychotherapist, Dr. Wei Lyu currently serves as the Vice President of Wenzhou Kangning Hospital and an adjunct professor at the School of Mental Health, Wenzhou Medical University; He has extensive practical experiences as a psychiatric practitioner. He is also actively involved in the field of mental health as a professional as well as a leader. He is also a Youth Member with the Society of Psychiatrists, Zhejiang Medical Doctor Association.
\end{IEEEbiography}

\begin{IEEEbiography}[{\includegraphics[width=1in,height=1.25in,clip,keepaspectratio]{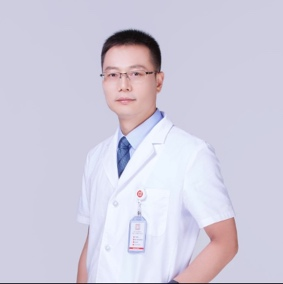}}]{Minjie Ye}
Dr. Minjie Ye is the Vice President and Director of the Center for Child and Adolescent Mental Health Services of the Affiliated Kangning Hospital of Wenzhou Medical University. He also serves as the Academic Lead for Kangning Hospital's National Key Clinical Specialty (Psychiatry). In addition to his position as the Associate Dean of the School of Mental Health at Wenzhou Medical University, Dr. Ye also heads the university's Child Mental Health Research Center. Dr. Ye is a member of the Society of Psychiatrists at Chinese Medical Doctor Association as well as a member of Committee of Marriage and Family Psychology with the Chinese Association of Social Psychology.
\end{IEEEbiography}

\begin{IEEEbiography}[{\includegraphics[width=1in,height=1.25in,clip,keepaspectratio]{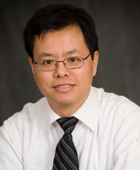}}]{Jie Chen}
Jie Chen received his Ph.D. degrees from the University of Maryland at College Park, USA. He is currently a Professor at the University of Alberta. Dr. Chen is an IEEE Fellow, a Fellow of the Canadian Academy of Engineering, a Fellow of American Institute of Medical and Biological Engineering. He received distinguished Alumni Award, the Department of Electrical and Computer Engineering, University of Maryland. He was also awarded Killam Annual Professorship, which is among the highest honours given to a professor at Canadian Universities.
\end{IEEEbiography}

\end{document}